\documentclass[journal]{IEEEtran}
\usepackage{epsfig,latexsym,graphics,psfrag,cite}
\usepackage[mathcal]{euscript}  

\def\ps@first{%
  \def\@oddhead{\hfil{\small \status}\hfil}%
  \def\@evenhead{\hfil{\small \status}\hfil}}

\usepackage{amssymb,amsmath,amsfonts,oldgerm,euscript}

\newtheorem{lemma}{Lemma}
\newtheorem{theorem}{Theorem}
\newtheorem{definition}{Definition}

\newcommand{\ful}[1]{#1_1^n}

\newcommand{\thrmref}[1]{Theorem~\mbox{\ref{#1}}}
\newcommand{\lemref}[1]{Lemma~\mbox{\ref{#1}}}

\newcommand{\secref}[1]{Section~\mbox{\ref{#1}}}

\newcommand{\defeq}{\stackrel{\Delta}{=}}

\newcommand{\qed}{\rule[0.1ex]{1.4ex}{1.6ex}}

\renewcommand{\ful}[1]{#1^n}
\newcommand{\comp}{\mathrm{c}}

\newcommand{\nSrc}{S}
\newcommand{\nsrc}{s}

\newcommand{\nChIn}{X}
\newcommand{\nchin}{x}

\newcommand{\nAux}{U}
\newcommand{\naux}{u}

\newcommand{\nAuxDeg}{U}
\newcommand{\nauxdeg}{u}

\newcommand{\nAuxRef}{T}
\newcommand{\nauxref}{t}

\newcommand{\nChOut}{Y}
\newcommand{\nchout}{y}

\newcommand{\nChOutRef}{\nChOut_{\mathrm{f}}}
\newcommand{\nchoutref}{\nchout_{\mathrm{f}}}

\newcommand{\nChOutDeg}{\nChOut_{\mathrm{c}}}
\newcommand{\nchoutdeg}{\nchout_{\mathrm{c}}}

\newcommand{\dtag}{\tau} 

\newcommand{\srcAlph}{\mathcal{S}}

\newcommand{\auxAlph}{\mathcal{U}}

\newcommand{\chinAlph}{\mathcal{X}}

\newcommand{\dsign}{\EuScript{S}}

\newcommand{\dver}{\EuScript{V}}

\newcommand{\Dist}{D}
\newcommand{\Diste}{D_{\mathrm{e}}}
\newcommand{\Distr}{D_{\mathrm{r}}}
\newcommand{\Distei}{D_{\mathrm{e},i}}
\newcommand{\Distri}{D_{\mathrm{r},i}}
\newcommand{\Distrc}{D_{\mathrm{r}}^{\mathrm{c}}}
\newcommand{\Distrf}{D_{\mathrm{r}}^{\mathrm{f}}}
\newcommand{\diste}{d_{\mathrm{e}}}
\newcommand{\distr}{d_{\mathrm{r}}}
\newcommand{\distrc}{d_{\mathrm{r}}}
\newcommand{\distrf}{d_{\mathrm{r}}}

\newcommand{\dfail}{\varnothing}

\newcommand{\encoder}{\Upsilon_n}
\newcommand{\secKey}{\theta}
\newcommand{\pubKey}{\secKey_p}
\newcommand{\privKey}{\secKey_s}

\newcommand{\decoder}{\Phi_n}
\newcommand{\xdecn}[1]{\Phi_{n}\left(#1\right)}
\newcommand{\xdec}[1]{\Phi\left(#1\right)}
\newcommand{\xdeci}[1]{\Phi_{i}\left(#1\right)}

\newcommand{\refsedec}[2]{g_{\mathrm{f}}\left(#1,#2\right)}
\newcommand{\degsedec}[1]{g_{\mathrm{c}}\left(#1\right)}

\newcommand{\encSNotTyp}{\mathcal{E}_\mathrm{st}}
\newcommand{\encFail}{\mathcal{E}_\mathrm{et}}
\newcommand{\encChFail}{\mathcal{E}_\mathrm{ct}}
\newcommand{\decFail}{\mathcal{E}_\mathrm{dt}}
\newcommand{\edfail}{\mathcal{E}_\mathrm{tf}}
\newcommand{\noedfail}{\edfail^\comp}

\newcommand{\exdiste}{\mathcal{E}_{\Diste}}
\newcommand{\exdistr}{\mathcal{E}_{\Distr}}
\newcommand{\Edv}{\mathcal{E}_\mathrm{dv}}

\newcommand{\undetErr}{\mathcal{E}_\mathrm{sa}}

\newcommand{\cbook}{\mathcal{C}}
\newcommand{\cbookdeg}{\mathcal{C}_{\mathrm{c}}} 
\newcommand{\cbookref}{\mathcal{C}_{\mathrm{f}}} 
\newcommand{\codeword}[1]{c_{#1}}

\newcommand{\admissC}{\mathcal{A}}
\newcommand{\cbkR}{R}
\newcommand{\cbkRdeg}{R_{\mathrm{c}}}
\newcommand{\cbkRref}{R_{\mathrm{f}}}
\newcommand{\cdeg}{c_{\mathrm{c}}}
\newcommand{\cdegh}{\hat{c}_{\mathrm{c}}}
\newcommand{\cref}{c_{\mathrm{f}}}
\newcommand{\crefh}{\hat{c}_{\mathrm{f}}}

\newcommand{\card}[1]{\left|#1\right|}
\newtheorem{prop}{Proposition}

\newcommand{\choutAlph}{\mathcal{Y}}

\newcommand{\sedec}[1]{g(#1)}
\newcommand{\sedeci}[2]{g_{#1}(#2)}
\newcommand{\fulsedec}[1]{\ful{g}(#1)}
\newcommand{\senc}[1]{f(#1)}

\newcommand{\reals}{\mathbb{R}}

\newcommand{\nIChOut}{Y}

\newcommand{\dmax}[1]{\bar{d}_{#1}}

\newcommand{\genrv}{T}

\newcommand{\nCode}{C}

\newcommand{\nSrch}{\hat{\nSrc}}
\newcommand{\nSrct}{\tilde{\nSrc}}
\newcommand{\nsrch}{\hat{\nsrc}}

\newcommand{\crossProb}{p}

\begin{document}

\title{Authentication with Distortion Criteria}
\author{Emin~Martinian,~\IEEEmembership{Member,~IEEE,} Gregory~W.~Wornell,~\IEEEmembership{Fellow,~IEEE} and~Brian~Chen~\IEEEmembership{Member,~IEEE}%
\thanks{Manuscript received May 2002; revised January 2004 and
  February 2005.  This work has been   supported in part by the National 
Science Foundation under Grant No.~CCR-0073520 and through a National
Science Foundation Graduate Fellowship, Microsoft Research,
Hewlett-Packard through the MIT/HP Alliance, and Texas Instruments
through the Leadership Universities Program.  This work was presented
in part at ISIT-2001, Washington, DC.}%
\thanks{The authors are affiliated with the Department of Electrical
Engineering and Computer Science, Massachusetts Institute of
Technology, Cambridge, MA 02139.   (E-mail: \{emin,gww,bchen\}@mit.edu).}}

\markboth{IEEE Trans.\ Inform.\ Theory,~Vol.~X, No.~XX,~~2005}{Martinian\MakeLowercase{\textit{et al.}}:
  Authentication with   Distortion Criteria}


\maketitle


\begin{abstract}
In a variety of applications, there is a need to authenticate content
that has experienced legitimate editing in addition to potential
tampering attacks.  We develop one formulation of this problem based
on a strict notion of security, and characterize and interpret the
associated information-theoretic performance limits.  The results can
be viewed as a natural generalization of classical approaches to
traditional authentication.  Additional insights into the structure of
such systems and their behavior are obtained by further specializing
the results to Bernoulli and Gaussian cases.  The associated systems
are shown to be substantially better in terms of performance and/or
security than commonly advocated approaches based on data hiding and
digital watermarking.  Finally, the formulation is extended to obtain
efficient layered authentication system constructions.
\end{abstract}


\section{Introduction}

\PARstart{I}{n} traditional authentication problems, the goal is to
determine whether some content being examined is an exact replica of
what was created by the author.  Digital signature techniques
\cite{diffie_hellman} are a natural tool for addressing such problems.
In such formulations, the focus on exactness avoids consideration of
semantic issues.  However, in many emerging applications, semantic
issues are an integral aspect of the problem, and cannot be treated
separably.  As contemporary examples, the content of interest may be
an audio or video waveform, or an image, and before being presented to
a decoder the waveform may experience any of a variety of possible
perturbations, including, for example, degradation due to noise or
compression; transformation by filtering, resampling, or transcoding;
or editing to annotate, enhance, or otherwise modify the waveform.
Moreover, such perturbations may be intentional or unintentional,
benign or malicious, and semantically significant or not.  Methods for
reliable authentication from such perturbed data are important as
well.

The spectrum of applications where such authentication capabilities
will be important is enormous, ranging from drivers' licenses,
passports, and other government-issued photo identication; to news
photographs and interview tapes; to state-issued currency and other
monetary instruments; to legal evidence in the form of audio and video
recordings in court cases.  Indeed, the rapidly increasing ease with
which such content can be digitally manipulated in sophisticated ways
using inexpensive systems, whether for legitimate or fraudulent
purposes, is of considerable concern in these applications.

Arising out of such concerns, a variety of technologies have been
introduced to facilitate authentication in such settings.  Examples
include various physical watermarking technologies --- such as
hologram imprinting in images --- as well as more recent digital
decendents.  See, e.g., \cite{pak99} for some of the rich history in
this area going back several hundred years.  However, regardless of
the implementation, all involve the process of marking or altering the
content in some way, which can be viewed as a form of encoding.

A rather generic problem that encompasses essentially all the
applications of interest is that of transaction-tracking in a content
migration scenario.  In this scenario, there are essentially three
types of participants involved in the migration of a particular piece
of content.  There is the original author or creator of the content,
who delivers an encoding of it.\footnote{There are no inherent
restrictions on what can constitute ``content'' in this generic
problem.  Typical examples include video, audio, imagery, text, and
various kinds data.}  There is the editor who makes modifications to
this encoded content, and publishes the result.\footnote{The motives
and behavior of the editor naturally depend on the particular
application and situation.  At one extreme the editor
might just perform some benign resampling or other transcoding, or, at
the other extreme, might attempt to create a forgery from the content.
In the latter case, the editor would be considered an attacker.}  And
there is the reader or end-user for whom the published work is
intended.  The reader wants to be able to determine 1) whether
published work being examined was derived from content originally
generated by the author, and 2) how it was modified by the editor.  At
the same time, the editor wants the author's encoding to be
(semantically) close to the original content, so that the
modifications can take the semantics into account as necessary.

In the recent literature, researchers have proposed a variety of
approaches to such problems based on elements of digital watermarking,
cryptography, and content classification; see, e.g., \cite{fridrich,
rey_2000, wolfgang_1996, friedman, kundur, wong, wu_liu, queluz,
bat_kut, md00, eggers_2001, yeung_1997, schneider_1996, Lin_2001,
Me_2001, Lu_2001} and the references therein.  Ultimately, the methods
developed to date implicitly or explicitly attempt to balance the
competing goals of robustness to benign perturbations, security
against tampering attacks, and encoding distortion.  

Within this literature, there are two basic types of approaches.  In
the first, the authentication mechanism is based on embedding what is
referred to as a ``fragile'' watermark known to both encoder and
decoder into the content of interest.  At the decoder, a watermark is
extracted and compared to the known watermark inserted by the encoder.
The difference between the extracted watermark and the known watermark
is then interpreted as a measure of authenticity.  Examples of this
basic approach include \cite{kundur, yeung_1997, wolfgang_1996,
eggers_2001}.

The second type of approach is based on a ``robust'' watermarking
strategy, whereby the important features of the content are extracted,
compressed and embedded back into the content by the encoder.  The
decoder attempts to extract the watermark from the content it obtains
and authenticates by comparing the features encoded in the watermark
to the features in the content itself.  This strategy is sometimes
termed ``self-embedding.''  Examples of this basic approach include
\cite{rey_2000, bat_kut, schneider_1996}.

Despite the growing number of proposed systems, many basic questions
remain about 1) how to best model the problem and what we mean by
authentication, 2) what the associated fundamental performance limits
are, and 3) what system structures can and cannot approach those
limits.  More generally, there are basic questions about the degree to
which the authentication, digital watermarking, and data hiding
problems are related or not.

While information-theoretic treatments of authentication problems are
just emerging, there has been a growing literature in the information
theory community on digital watermarking and data hiding problems, and
more generally problems of coding with side information, much of which
builds on the foundation of \cite{gelfand_1980, costa_83, heg83}; see,
e.g., \cite{mos98, cw00b, cl00, mos00, chen_wornell_2001, moulin2003,
sm01, cohen_2002, swanson, memon, cox, cpr99, pcr03, seg00,
rjb_bc_gw_preprint, bcw01, Merhav_2000, cc01, esz00, zse02,
Sutivong_2002} and the references therein.  Collectively, this work
provides a useful context within which to examine the topic of
authentication.

Our contribution in this paper is to propose one possible formulation
for the general problem of authentication with a semantic model, and
examine its implications.  In particular, using distortion criteria to
capture semantic aspects of the problem, we assess performance limits
in terms of the inherent trade-offs between security, robustness, and
distortion, and in turn develop the structure of systems that make
these trade-offs efficiently.  As we will show, these systems have
important distinguishing characteristics from those proposed to date.
We also see that under this model, the general authentication problem
is substantially different from familiar formulations of the digital
watermarking and data hiding problems, and has a correspondingly
different solution.

A detailed outline of the paper is as follows.  We begin by briefly
defining our notation and terminology in \secref{sec:notation}.  Next
in \secref{sec:informal_problem}, we develop a system model and
problem formulation, quantifying a notion of authentication.  In
\secref{sec:codethms}, we characterize the performance limits of such
systems via our main coding theorem.  \secref{sec:proofs} contains
both the associated achievability proof, which identifies the
structure of good systems, and a converse.  In
\secref{sec:binary_hamming} the results are applied to the case of
binary content with Hamming distortion measures, and in
\secref{sec:gaussian} to Gaussian content with quadratic distortion
measures.  \secref{sec:discussion} then analyzes other classes of
authentication techniques in the context of our framework, and shows
that they are inherently either less efficient or less secure that the
systems developed here.  Next, \secref{sec:layered} generalizes the
results of the paper to include layered systems that support multiple
levels of authentication.  Finally, \secref{sec:conc} contains some
concluding remarks.

\section{Notation and Terminology}
\label{sec:notation}

We use standard information theory notation (e.g., as found in
\cite{cover}).  Specifically, $E[A]$ denotes expectation of the random
variable $A$, $H(A)$, and $I(B;C)$ denote entropy and mutual
information, and $A \leftrightarrow B \leftrightarrow C$ denotes the
Markov condition that random variables $A$ and $C$ are independent
given $B$.  We use the notation $v_i^j$ to denote the sequence
$\{v_i,v_{i+1},\dots,v_j\}$, and define $\ful{v}=v_1^n$.  Alphabets
are denoted by uppercase calligraphic letters, e.g., $\srcAlph$,
$\chinAlph$.  We use $\card{\cdot}$ to denote the cardinality of a set
or alphabet.

Since the applications are quite varied, we keep our terminology
rather generic.  The content of interest, as well as its various
encodings and recontructions, will be generically referred to as
``signals,'' regardless of whether they refer to video, audio,
imagery, text, data, or any other kind of content.  The original
content we will also sometimes simply refer to as the ``source.''
Moreover, we will generally associate any manipulations of the encoded
content with the ``editor,'' regardless of whether any human is
involved.  However, as an exception, we will often use the term
``attacker'' in lieu of ``editor'' for cases where the manipulations
are specifically of a malicious nature.

\section{System Model and Problem Formulation}
\label{sec:informal_problem}

\begin{figure*}[tbp]
\centering
\psfrag{S}{\huge$\ful{\nSrc}$}
\psfrag{X}{\huge$\ful{\nChIn}$}
\psfrag{Y}{\huge$\ful{\nChOut}$}
\psfrag{Sh}{\huge$\ful{\nSrch}$ or $\dfail$}
\includegraphics[angle=0,width=5in]{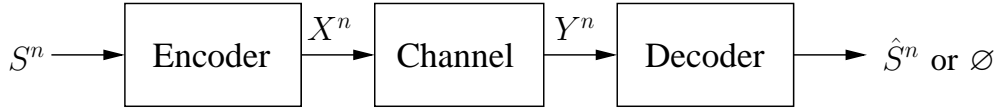}
\caption{Authentication system model.  The source $\ful{\nSrc}$ is
encoded by the content creator into $\ful{\nChIn}$, incurring some
distortion.  The channel models the actions of the editor, i.e., all
processing experienced by the encoded content before it is made
available to the end-user.  The decoder, controlled by the end-user,
produces from the channel output $\ful{\nChOut}$ either an authentic
reconstruction $\ful{\nSrch}$ of the source to within some fidelity,
or indicates that authentication is not possible using the special
symbol $\dfail$.
\label{fig:channel}}
\end{figure*}

Our system model for the transaction-tracking scenario is as depicted
in Fig.~\ref{fig:channel}.  To simplify the exposition, we model the
original content as an independent and identically distributed
(i.i.d.)\footnote{Our results do not depend critically on the i.i.d.\
property, which is chosen for convenience.  In fact, the i.i.d.\ model
is sometimes pessimistic; better performance can often be obtained by
taking advantage of correlation present in the source or channel.  We
believe that qualitatively similar results would be obtained in more
general settings (e.g., using techniques from \cite{Verdu_1994,
Steinberg_1996}).}  sequence $\nSrc_1, \nSrc_2, \ldots, \nSrc_n$.  In
practice $\ful{\nSrc}$ could correspond to sample values or signal
representations in some suitable basis.

The encoder takes as input the block of $n$ source samples
$\ful{\nSrc}$, producing an output $\ful{\nChIn}$ that is suitably
close to $\ful{\nSrc}$ with respect to some distortion measure.  The
encoder is under the control of the content creator.  The encoded
signal then passes through a channel, which models the actions of the
generic ``editor'', and encompasses all processing experienced by the
encoded signal before it is made available to the end-user as
$\ful{\nChOut}$.  This processing would include all effects ranging
from routine handling to malicious tampering.  The decoder, which is
controlled by the end-user, either produces, to within some fidelity
as quantified by a suitable distortion measure, a reconstruction
$\ful{\nSrch}$ of the source that is guaranteed to be free from the
effects of any modifications by the editor, or declares that it is not
possible to produce such a reconstruction.  We term such
reconstructions ``authentic.''

Our approach to the associated channel modeling issues in the
formulation of Fig.~\ref{fig:channel} has some novel features, and
thus warrants special discussion.  Indeed, as we now discuss, our
approach to such modeling is not to \emph{anticipate} the possible
behaviors of the editor, but to effectively \emph{constrain} them.  In
particular, we avoid choosing a model that tries to characterize the
range of processing the editor might undertake.  If we did, the
security properties of the resulting system would end up being
sensitive to any modeling errors, i.e., to any behavior of the editor
that is inconsistent with the model.

Instead, the focus is on choosing a model that defines the range of
processing the editor can undertake and have such edits accepted by
the end-user.  We refer to this as our ``reference channel model.''
Specifically, we effectively design the system such the decoder will
successfully authenticate the modified content if and only if the
edits are consistent with the reference channel model.  Thus, the
editor is free to edit the content in any way (and we make no attempt
to model the range of behavior), but the subset of behaviors for which
the system will authenticate is strictly controlled via the reference
channel construct.  Ultimately, since the end-user will not accept
content that cannot be authenticated, the editor will constrain its
behavior according to the reference channel.  

From this perspective, the reference channel model is a system design
parameter, and thus is known a priori to encoders, decoders, and
editors.  To simplify our analysis, we will restrict our attention to
memoryless probabilistic reference channel models.  In this case, the
model is characterized by a simple conditional distribution
$p(\nChOut|\nChIn)$.

As our main result, in Section~\ref{sec:codethms} we characterize when
authentication systems with the above-described behavior are possible,
and when they are not.  Specifically, let $\Diste$ denote the encoding
distortion, i.e., the distortion experienced in the absence of a
channel, and let $\Distr$ denote the distortion in the reconstruction
produced by the decoder when the signal can be authenticated, i.e.,
when the channel transformations are consistent with the chosen
reference distribution $p(\nchout|\nchin)$.  Then we determine which
distortion pairs $(\Diste,\Distr)$ are asymptotically achievable.

We emphasize that the distortion pair $(\Diste,\Distr)$ corresponds
precisely to the performance characteristics of direct interest in the
system for the transaction-tracking scenario.  Indeed, a small
$\Diste$ means the editor is given work with a faithful version of the
original content.  Moreover, a small $\Distr$ means that the end-user
is able to accurately estimate the editor's modifications by comparing
the decoder input to the authentic reconstruction.

\subsection{Defining ``Authenticity''}

To develop our main results, we first need to quantify the concept of
an ``authentic reconstruction.''  Recall that our intuitive notion of
an authentic reconstruction is one that is free from the effects of
the edits when the reference channel is in effect.  Formally, this is
naturally expressed as follows.
\begin{definition} \label{def:authrec}
A reconstruction $\ful{\nSrch}$ produced by the decoder from the
output $\ful{\nChOut}$ of the reference channel is said to be
authentic if it satisfies the Markov condition below:
\begin{equation}
\ful{\nSrch} \leftrightarrow \{ \ful{\nSrc}, \ful{\nChIn} \}
\leftrightarrow \ful{\nChOut}
\label{eq:estmarkov}
\end{equation}
\end{definition}
Note that as special cases, this definition would include systems in
which, for example, $\ful{\nSrch}$ is a deterministic or randomized
function of $\ful{\nSrc}$.  More generally, this definition means that
the authentic reconstructions are effectively defined by the encoder
in such systems.  This will have implications later in the system
design.   

\subsection{An Example Distortion Region}
\label{sec:exdr}

Before developing our main result, we illustrate with an example the
kinds of results that will be obtained.  This example corresponds to a
problem involving a symmetric Bernoulli source, Hamming distortion
measures, and a (memoryless) binary symmetric reference channel with
crossover probability $p$.

Under this example scenario, the editor is allowed to flip a fraction
$p$ of the binary source samples, and the end-user must (almost
certainly) be able to generate an authentic reconstruction from such a
perturbation.  If the edits are generated from a different
distribution, such as a binary symmetric channel with a cross-over
probability greater than $p$, then the decoder must (almost certainly)
declare an authentication failure.

\begin{figure}[tbp]
\centering
\psfrag{&2}{\huge$\Diste$}
\psfrag{&1}{\huge$\Distr$}
\psfrag{&3}{\LARGE$p$}
\psfrag{&4}{\LARGE$\frac{1}{2}$}
\psfrag{&5}{\LARGE$p$}
\psfrag{&6}{\LARGE$\frac{1}{2}$}
\includegraphics[angle=0,width=3.5in]{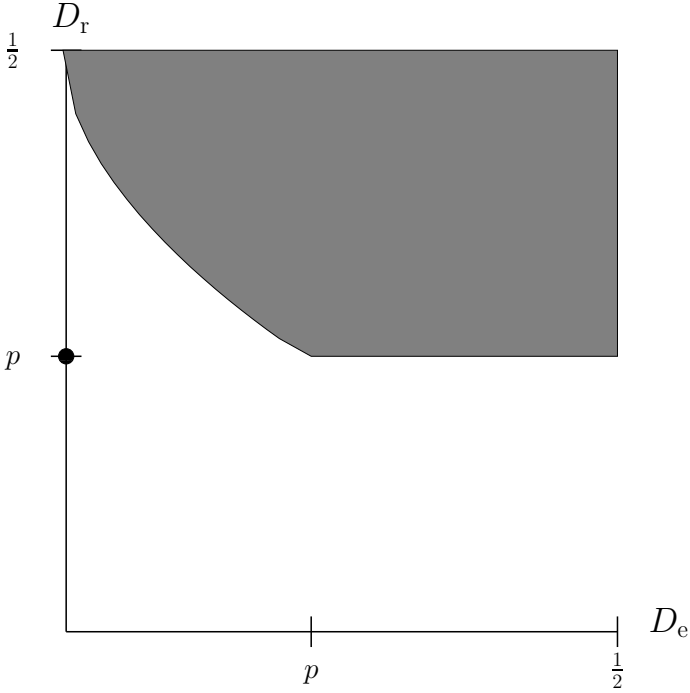}
\caption{The shaded area depicts the achievable distortion region for
a symmetric Bernoulli source used in conjunction with a binary
symmetric reference channel of crossover probability $p$.  Distortions
are with respect to the Hamming measure.  The case $p=0$ corresponds
to traditional digital signatures.  If authentication was not
required, the point $(\Diste = 0, \Distr = p)$ could be achieved.
\label{fig:ham_reg}}
\end{figure}

The corresponding achievable distortion region is depicted in
Fig.~\ref{fig:ham_reg}.  Several points on the frontier are worth
discussing.  First, note that the upper left point on the frontier,
i.e., $(\Diste,\Distr) = (0,1/2)$, reflects that if no encoding
distortion is allowed, then authentic reconstructions are not
possible, since the maximum possible distortion is incurred.  At the
other extreme, the lower right point of the frontier, i.e.,
$(\Diste,\Distr) = (1/2,p)$, corresponds to a system in which the
source is first source coded to distortion $p$, afterwhich the
resulting bits are digitally signed and channel coded for the BSC.

While no amount of encoding distortion can reduce the reconstruction
distortion below $p$, the point $(\Diste,\Distr) = (p,p)$ on the
frontier establishes that a reconstruction distortion of $p$ is
actually achievable with much less encoding distortion than the lower
right point suggests.  In fact, because the required encoding
distortion is only $p$, the decoder can be viewed as completely
eliminating the effects of the reference channel when it is in effect:
the minimum achievable reconstruction distortion $\Distr$ is the same
as the distortion $\Diste$ at the output of the encoder.

The more general structure of the frontier is also worth observing.
In particular, $\Distr$ is a decreasing function of $\Diste$ along the
frontier.  This reflects that the objectives of small $\Diste$ (which
the editor wants) and a small $\Distr$ (which the end-user wants) are
conflicting and a fundamental tradeoff is involved for any given
reference channel.  In fact, as we will see in the sequel, this
behavior is not specific to this example, but a more general feature
of our authentication problem formulation.\footnote{This should not be
surprising, since such tradeoffs frequently arise in joint
source-channel coding problems with uncertain channels; see, e.g.,
\cite{Mittal_2002, Reznic_2002, Shamai_1998}.}

Finally, observe that the achievable region decreases monotonically
with $p$, the severity of edits allowed.  Thus, if one has particular
target encoding and reconstruction distortions, then this effectively
limits how much editing can be tolerated.  As the extreme point, the
case $p=0$ in which no editing is allowed corresponds to the
traditional scenario for digital signatures.  In this case, as the
figure reflects, authentication is achievable without incurring any
encoding distortion nor reconstruction distortion.  It is worth noting
that the nature of the interplay between the severity of the reference
channel and the achievable distortion region is not specific to this
example, but arises more generally with this formulation of the
authentication problem.

\section{Characterization of Solution: Coding Theorems}
\label{sec:codethms}

An instance of the authentication problem consists of the seven-tuple
\begin{equation} 
\left\{ \srcAlph, p(\nsrc), \chinAlph, \choutAlph, p(\nchout|\nchin), 
        \diste(\cdot,\cdot), \distr(\cdot,\cdot) \right\}.
\label{eq:authprob}
\end{equation}
We use $\srcAlph$ to denote the source alphabet---which is finite unless
otherwise indicated---and $p(\nsrc)$ is its (i.i.d.) distribution.  The
channel input and output alphabets are $\chinAlph$ and $\choutAlph$
and $p(\nchout|\nchin)$ is the (memoryless) reference channel law.
Finally, $\diste(\cdot,\cdot)$ and $\distr(\cdot,\cdot)$ are the
encoding and reconstruction distortion measures.

A solution to this problem (i.e., an authentication scheme) consists
of an algorithm that returns an encoding function $\encoder$, a
decoding function $\decoder$, and a secret key $\secKey$.  The secret
key is shared only between the encoder and decoder; all other
information is known to all parties including editors.  (For the
interested reader, straightforward adaptations of our solutions to
public-key implementations are summarized in the Appendix.  However,
we otherwise restrict our attention to private-key schemes in the
paper to focus the exposition.)

The secret key $\secKey$ is a $k$-bit sequence with $k$ sufficiently
large.  The encoder is a mapping from the source sequence and the
secret key to codewords, i.e.,
\begin{equation*}
\encoder(\ful{\nSrc},\secKey):\quad\srcAlph^n
\times \{0,1\}^k \mapsto \chinAlph^n.
\end{equation*}

The decoder is a mapping from the channel output and the secret key to
either an authentic source reconstruction $\ful{\nSrch}$ (i.e., one
satisfying \eqref{eq:estmarkov}) or the special symbol $\dfail$ that
indicates  such a reconstruction 
is not possible; whence,
\begin{equation*}
\xdecn{\ful{\nIChOut},\secKey}:\quad \choutAlph^n \times
\{0,1\}^k \mapsto  \srcAlph^n \cup \{\dfail\}.
\end{equation*}
Notice that since an authentic reconstruction must satisfy
\eqref{eq:estmarkov}, and since the decoder must satisfy the Markov
condition $\{\ful{\nSrc},\ful{\nChIn}\} \leftrightarrow \ful{\nChOut}
\leftrightarrow \xdecn{\ful{\nChOut},\secKey}$, we have that
$\ful{\nSrch} \leftrightarrow \{\ful{\nSrc},\ful{\nChIn}\}
\leftrightarrow \xdecn{\ful{\nChOut},\secKey}$ forms a Markov chain
only \emph{when successful decoding occurs}.  Thus, the
authentic reconstruction $\ful{\nSrch}$ should be defined as a
quantity that the decoder attempts to deduce since defining
$\ful{\nSrch} = \xdecn{\ful{\nChOut,\secKey}}$ will generally not
satisfy \eqref{eq:estmarkov}.

Henceforth, except when there is risk of confusion, we omit both the
subscript $n$ and the secret key argument from the encoding and
decoding function notation, letting the dependence be implicit.
Moreover, when the encoder and/or decoder are randomized functions,
then all probabilities are taken over these randomizations as well as
the source and channel law.

The relevant distortions are the encoding and decoding
distortion computed as the sum of the respective (bounded) single
letter distortion functions $\diste$ and $\distr$, i.e., 
\begin{equation*} 
\frac{1}{n} \sum_{i=1}^n \diste(\nSrc_i,\nChIn_i)\qquad\text{and}\qquad
\frac{1}{n} \sum_{i=1}^n \distr(\nSrc_i,\xdeci{\ful{\nChOut}}).
\end{equation*}
Evidently,
\begin{align}
\diste &:\quad \srcAlph\times\chinAlph \mapsto \reals^+  \\
\distr &:\quad \srcAlph\times\srcAlph \mapsto \reals^+.
\end{align}

The system can fail in one of three ways.  The first two failure modes
correspond to either the encoder introducing excessive encoding
distortion, or the decoder failing to produce an authentic
reconstruction with acceptable distortion when the reference channel
is in effect.  Accordingly, we define the overall distortion violation
error event to be
\begin{equation} 
\Edv = \exdiste \cup \exdistr
\label{eq:Edv-def}
\end{equation}
where, for any $\epsilon>0$,
\begin{align}
\exdiste
&= \left\{\frac{1}{n}\sum_{i=1}^n \diste(\nSrc_i,\nChIn_i) 
    > \Diste +\epsilon \right\} \label{eq:d1def}\\
\exdistr
&= \bigg\{\xdecn{\ful{\nChOut}} = \dfail \bigg\} \notag\\
& \ \ \ \ \cup \left\{ \frac{1}{n}\sum_{i=1}^n
\distr(\nSrc_i,\xdeci{\ful{\nChOut}}) 
    > \Distr + \epsilon \right\} \notag\\
& \ \ \ \ \cap \bigg\{\xdecn{\ful{\nChOut}} \neq \dfail \bigg\}.
    \label{eq:d2def} 
\end{align}

In the remaining failure mode, the system fails to produce the desired
authentic reconstruction $\ful{\nSrch}$ from the channel output and
instead of declaring that authentication is not possible produces an
incorrect estimate.  Specifically, we define the successful attack
event according to
\begin{equation} 
\undetErr = 
\{ \xdec{\ful{\nIChOut}} \neq \dfail \} \cap 
\{ \xdec{\ful{\nIChOut}} \neq \nSrch^n \}. 
\label{eq:undetErr-def}
\end{equation}


\begin{definition}
\label{def:adr}
The achievable distortion region for the problem \eqref{eq:authprob}
is the closure of the set of pairs $(\Diste,\Distr)$ such that there
exists a sequence of authentication systems, indexed by $n$, where for
every $\epsilon > 0$ and as $n\rightarrow\infty$,
$\Pr[\undetErr]\rightarrow0$ regardless of the channel law in effect,
$\Pr[\exdiste]\rightarrow0$, and $\Pr[\exdistr]\rightarrow0$
when the reference channel is in effect, with $\undetErr$, $\exdiste$,
and $\exdistr$ as defined in \eqref{eq:undetErr-def},
\eqref{eq:d1def}, and \eqref{eq:d2def}.
\end{definition}

For such systems, we have the following coding theorem:
\begin{theorem}
\label{th:main}
The distortion pair $(\Diste,\Distr)$ lies in the achievable
distortion region for the problem \eqref{eq:authprob} if and only if
there exist functions $\senc{\cdot,\cdot}$, $\sedec{\cdot}$ and 
a distribution $p(\nchout,\nchin,\naux,\nsrc) =
p(\nsrc)p(\naux|\nsrc)p(\nchin|\naux,\nsrc)p(\nchout|\nchin)$ with
$\nChIn$ deterministic
(i.e. $p(\nchin|\naux,\nsrc)=1_{\nchin=\senc{\nsrc,\naux}}$) 
such that
\begin{subequations}
\label{eq:thm}
\begin{align}
I(\nAux;\nChOut) - I(\nSrc;\nAux)  &\geq  0 \label{eq:thm:a} \\
E[\diste(\nSrc,\senc{\nAux,\nSrc})]  &\leq  \Diste \label{eq:thm:b} \\
E[\distr(\nSrc,\sedec{\nAux})]  &\leq  \Distr. \label{eq:thm:c}
\end{align}
The alphabet $\auxAlph$ of the auxiliary random variable $\nAux$
requires cardinality $\card{\auxAlph}
\le (\card{\srcAlph} + \card{\chinAlph} +
3)\cdot\card{\srcAlph}\cdot\card{\chinAlph}$.\footnote{\textnormal{If
instead $f(\nAux,\nSrc)$ is allowed to be a non-deterministic mapping,
then it is sufficient to consider distributions where the auxiliary
random variable has the smaller alphabet $\card{\auxAlph} \le
\card{\srcAlph} + \card{\chinAlph} + 3$.}}
\end{subequations}
\end{theorem}

Essentially, the auxiliary random variable $\nAux$ represents an
embedded description of the source that can be authenticated, $\nChIn$
represents the encoding of the source $\nSrc$, and $\sedec{\nAux}$ in
\eqref{eq:thm:c} represents the authentic reconstruction.  The usual
condition that the channel output is determined from the channel input
(i.e., the encoder does not know what the channel output will be until
after the channel input is fixed) is captured by the requirement that
the full joint distribution $p(\nchout,\nchin,\naux,\nsrc)$ factors as
shown above.  The requirement \eqref{eq:estmarkov} that the authentic
reconstruction does not depend directly on the editors manipulations
--- i.e., the realization of the reference channel --- is captured by
the fact that $\sedec{\cdot}$ depends only on $\nAux$ and not on
$\nChOut$.  Without the authentication requirement, the set of
achievable distortion pairs can be enlarged by allowing the
reconstruction to depend on the channel output, i.e.\ $\sedec{\nAux}$
in \eqref{eq:thm:c} can be replaced by $\sedec{\nAux,\nChOut}$.  Thus,
as we shall see in Sections~\ref{sec:binary_hamming} and
\ref{sec:gaussian}, security comes at a price in this problem.

Theorem~\ref{th:main} has some interesting features.  First, it is
worth noting that since the problem formulation is inherently
``analog,'' dealing only with waveforms, we might expect the best
solutions to the problem to be analog in nature.  However, what the
theorem suggests, and what its proof confirms, is that digital
solutions are in fact sufficient to achieve optimality.  In
particular, as we will see, source and channel coding based on
discrete codebooks are key ingredients of the achievability argument.
In some sense, this is the consequence of the inherently discrete
functionality we have required of the decoder with our formulation.

As a second remark, note that Theorem~\ref{th:main} can be contrasted
with its information embedding counterpart, which as generalized from
\cite{gelfand_1980} in \cite{rjb_bc_gw_preprint}, states that a pair
$(R,\Diste)$, where $R$ is the embedding rate, is achievable if and
only if there exists a function $\senc{\cdot,\cdot}$ and a
distribution $p(\nchout,\nchin,\naux,\nsrc) =
p(\nsrc)p(\naux|\nsrc)p(\nchin|\nsrc,\naux)p(\nchout|\nchin)$ with
$\nChIn$ deterministic
(i.e. $p(\nchin|\naux,\nsrc)=1_{\nchin=\senc{\nsrc,\naux}}$) such that
\begin{subequations}
\label{eq:ie-thm}
\begin{align}
I(\nAux;\nChOut) - I(\nSrc;\nAux)  &\geq  R \label{eq:ie:a} \\
E[\diste(\nSrc,\senc{\nAux,\nSrc})]  &\leq  \Diste. \label{eq:ie:b}
\end{align}
Thus we see that the authentication problem is substantially
different from the information embedding problem.
\end{subequations}

Before developing the proofs of Theorem~\ref{th:main}, to develop
intuition we describe the general system structure, and its
specialization to the Gaussian-quadratic case.

\subsection{General System Structure}
\label{sec:geometric_view}

As developed in detail in \secref{sec:proofs}, an optimal
authentication system can be constructed by choosing a codebook
$\cbook$ with codewords appropriately distributed over the space of
possible source outcomes.  The elements of a randomly chosen subset of
these codewords $\admissC \subset \cbook$ are marked as admissible and
the knowledge of $\admissC$ is a secret shared between the encoder and
decoder, and kept from editors.

The encoder maps (quantizes) the source $\ful{\nSrc}$ to the nearest
admissible codeword $\ful{\nAux}$ and then generates the channel input
$\ful{\nChIn}$ from $\ful{\nAux}$.  The decoder maps the signal it
obtains to the nearest codeword $\ful{\nCode}\in\cbook$.  If
$\ful{\nCode}\in\admissC$, i.e., $\ful{\nCode}$ is an admissible
codeword, the decoder produces the reconstruction $\ful{\nSrch}$ from
$\ful{\nCode}$.  If $\ful{\nCode}\not\in\admissC$, i.e.,
$\ful{\nCode}$ is not admissible, the decoder declares that an
authentic reconstruction is not possible.

Observe that the $\admissC$ must have the following three
characteristics.  First, to avoid a successful attack the number of
admissible codewords must be appropriately small.  Indeed, since
attackers do not know $\admissC$, if an attacker's tampering causes
the decoder to decode to any codeword other than $\ful{\nAux}$ then
the probability that the decoder is fooled by the tampering and does
not declare a decoding failure is bounded by
$\card{\admissC}/\card{\cbook}$.  Second, to avoid an encoding
distortion violation, the set of admissible codewords should be dense
enough to allow the encoder to find an appropriate $\ful{\nChIn}$ near
$\ful{\nSrc}$.  Third, to avoid a reconstruction distortion violation,
the decoder should be able to distinguish the possible encoded signals
at the output of the reference channel.  Thus the codewords should be
sufficiently separated that they can be resolved at the output of the
reference channel.

\subsubsection{Geometry for Gaussian-Quadratic Example}
\label{sec:sphere_packing}

\begin{figure*}[tbp]
\centering
\epsfbox{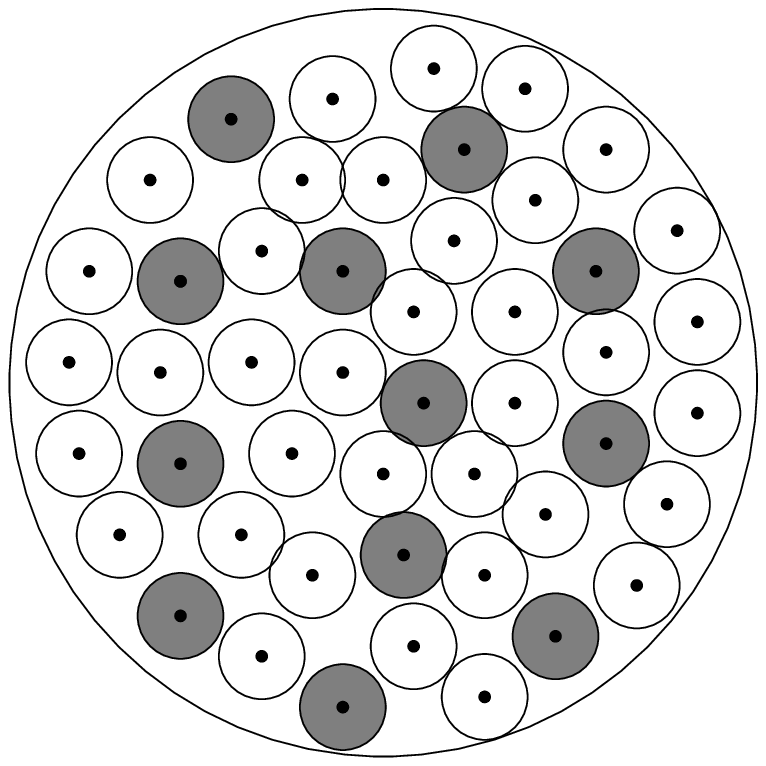}
\caption{Codebook construction for the Gaussian-quadratic scenario.
The large sphere represents the space of possible source vectors and
the small spheres representing the noise are centered on codewords.
When the small spheres do not overlap, the codewords can be resolved at
the output of the reference channel.  The shaded spheres represent the
admissible codewords---a secret known only to the encoder and decoder.
\label{fig:sphere_packing}}
\end{figure*}

We illustrate the system geometry in the case of a white Gaussian
source, quadratic distortion measure, and an additive white Gaussian
noise reference channel, in the high signal-to-noise ratio (SNR)
regime.  We let $\sigma_{\nSrc}^2$ and $\sigma_N^2$ denote the source
and channel variances, respectively.  For this example, we can
construct $\cbook$ by packing codewords into the space of possible
source vectors such that no codeword is closer than some distance
$r\sqrt{n}$ to any other, i.e., packing spheres of radius $r\sqrt{n}$
into a sphere of radius $\sigma_{\nSrc}\sqrt{n}$ where the center of
the spheres correspond to codewords.  Next, a fraction $2^{-n\gamma}$
of the codewords in $\cbook$ are chosen at random and marked as
admissible to form $\admissC$.  It suffices to let $\gamma=1/\sqrt{n}$
and $r^2=\sigma_N^2+\epsilon$ for some $\epsilon>0$ that is
arbitrarily small.  This construction is illustrated in
Fig.~\ref{fig:sphere_packing}.

The encoder maps the source $\ful{\nSrc}$ to a nearby admissible
codeword $\ful{\nAux}$, which it chooses as the encoding
$\ful{\nChIn}$.  Since the number of admissible codewords in a sphere
of radius $d$ centered on $\ful{\nSrc}$ is roughly
\begin{equation*} 
\frac{\card{\admissC}}{\card{\cbook}} \cdot \left(\frac{d}{r}\right)^n,
\end{equation*}
on average there exists at least one codeword within distance $d$ of
the source provided $d \geq r 2^{\gamma}$.  Thus, the average
encoding distortion is roughly $r^2 2^{2\gamma}$, which approaches
$\sigma_N^2+\epsilon$ as $n\rightarrow\infty$.

The authentic reconstruction is $\ful{\nSrch} = \ful{\nAux}$.  Thus,
when the decoder correctly identifies $\ful{\nAux}$, the
reconstruction distortion is the same as the encoding distortion.  And
when the reference channel is in effect, the decoder does indeed
correctly identify $\ful{\nAux}$.  This follows from the fact that
with high probability, the reference channel noise creates a
perturbation within a noise sphere of radius $\sigma_N \sqrt{n}$ about
the encoding $\ful{\nChIn}$, and the noise spheres do not
intersect since $r>\sigma_N$.

Furthermore, when the reference channel is not in effect and an
attacker tampers with the signal such that the nearest codeword
$\nCode$ is different from that chosen by the encoder $\ful{\nAux}$,
then the probability that $\nCode$ was marked as admissible in the
codebook construction phase is
\begin{equation*} 
\Pr[\nCode \in \admissC| \nCode \neq \ful{\nAux}] =
\frac{\card{\admissC}}{\card{\cbook}} = 2^{-n\gamma},
\end{equation*}
which goes to zero as $n\rightarrow\infty$.  The decoder generates
$\dfail$ if it decodes to a non-admissible codeword, so the
probability of a nonauthentic reconstruction is vanishingly small.

Thus the distortions $\Diste=\Distr=\sigma_N^2$ can be approached with
an arbitrarily small probability of successful attack.  See
\cite{mthesis, martinian_2001} for insights into the
practical implementation of this class of systems including those
designed based on a public key instead of a secret key.

\section{Proofs}
\label{sec:proofs}

\subsection{Forward Part: Sufficiency}
\label{sec:forw-part:-suff}

Here we show that if there exist distributions and functions
satisfying \eqref{eq:thm}, then for every $\epsilon >0 $ there exists
a sequence of authentication system with distortion at most
$(\Diste+\epsilon,\Distr+\epsilon)$.  Since the achievable distortion region
is a closed set this implies that $(\Diste,\Distr)$ lies in the
achievable distortion region.

We prove this forward part of \thrmref{th:main} by showing the
existence of a random code with the desired properties.  

\subsubsection{Codebook Generation}

We begin by choosing some $\gamma>0$ such that
\begin{equation} 
I(\nChOut;\nAux) - I(\nAux;\nSrc) > 3\gamma.
\label{eq:gammadef}
\end{equation}
where $\gamma$ decays to zero more slowly than $1/n$, i.e.,
\begin{equation} 
\text{$\gamma\rightarrow0$ and $n\gamma\rightarrow\infty$ as 
$n\rightarrow\infty$}.
\label{eq:gamma-props}
\end{equation}
Given the choice of $\gamma$, the encoder chooses a random
codebook $\cbook$ of rate
\begin{equation} 
\cbkR = I(\nSrc;\nAux) + 2\gamma.  
\label{eq:Rdef}
\end{equation}
Each codeword in $\cbook$ is a
sequence of $2^{n\cbkR}$ i.i.d.\ random variables selected according
to the distribution $p(\naux) = \sum_{\nsrc \in \srcAlph}\,
p(\naux | \nsrc ) p( \nsrc )$.  
Then, for each realized codebook $\cbook$ the encoder randomly marks 
$2^{n(\cbkR-\gamma)}$ of the codewords in $\cbook$ as
admissible and the others as forbidden.  We denote this new codebook of
admissible codewords as $\admissC$, which has effective rate
\begin{equation} 
\cbkR' = \cbkR - \gamma = I(\nSrc;\nAux) + \gamma,
\label{eq:Rpdef}
\end{equation}
where the last equality follows from substituting \eqref{eq:Rdef}.
The knowledge of which codewords are forbidden is the secret key and
is revealed only to the decoder.  The codebook $\cbook$ is publicly
revealed.

\subsubsection{Encoding and Decoding}

The encoder first tries to find an admissible codeword $\ful{\naux}
\in \admissC$ that is $\delta$-strongly jointly typical with its
source sequence $\ful{\nSrc}$ according to $p(\naux|\nsrc)$.  If the
codeword $\ful{\naux} \in \admissC$ is found to be typical, the
encoder output is produced by mapping the pair
$(\ful{\nsrc},\ful{\naux})$ into $\ful{\nchin}$ via
$\nchin=f(\nsrc,\naux)$.  If no jointly typical admissible codeword
exists, the encoder expects the system to fail, and thus selects an
arbitrary codeword.

The decoder attempts to produce the authentic reconstruction 
$\ful{\nsrch} = \fulsedec{\ful{\naux}}$ where
\begin{equation} 
\fulsedec{\ful{\naux}} = 
(\sedec{\naux_1}, \sedec{\naux_2}, \dots, 
 \sedec{\naux_n}).
\end{equation}
The decoder $\xdec{\cdot}$ tries to deduce $\ful{\nsrch}$
by searching for a unique admissible codeword
$\ful{\hat{\naux}} \in \admissC$ that is $\delta$-strongly jointly
typical with the obtained sequence $\ful{\nChOut}$.  If such a
codeword is found the reconstruction produced is
$\fulsedec{\ful{\hat{\naux}}}$.  If no such unique
codeword is found, the 
decoder produces the output symbol $\dfail$.  

\subsubsection{System Failure Probabilities}

We begin by analyzing the system failure probabilities.

\paragraph{Probability of Successful Attack.}

Suppose the attacker causes the codeword obtained by the decoder to be
jointly typical with a unique codeword $\ful{c}\in\cbook$.  Since the
attacker has no knowledge of which codewords are admissible, the
probability that codeword $\ful{c}$ was chosen as admissible in the
codebook construction phase is
\begin{equation*}
\Pr[\ful{c} \in \admissC] 
= \frac{\left|\admissC\right|}{\left|\cbook\right|} =
\frac{2^{n\cbkR'}}{2^{n\cbkR}} = 2^{-n\gamma}.
\end{equation*}
where we have used \eqref{eq:Rpdef} and \eqref{eq:Rdef}.  Therefore,
\begin{equation*}
\Pr[\undetErr] \leq \Pr[\xdec{\ful{\nIChOut}} \neq
\dfail \mid \xdec{\ful{\nIChOut}} \neq \ful{\nSrch}] = 2^{-n\gamma}.
\end{equation*}
which goes to zero according to \eqref{eq:gamma-props}.  Note that
this argument applies regardless of the method used by the attacker
since without access to the secret key its actions are statistically
independent of which codewords are admissible.

\paragraph{Probability of Distortion Violation.}

The distortion violation events $\exdiste$ and $\exdistr$ defined in
\eqref{eq:d1def} and \eqref{eq:d2def} can arise due to any of the
following typicality failure events:
\begin{itemize}
\item $\encSNotTyp$: The source is not typical.
\item $\encFail$: The encoder fails to find an admissible codeword
that is jointly typical with its input.
\item $\encChFail$: The channel fails to produce an output jointly
typical with its input when the reference channel law is in effect.
\item $\decFail$: The decoder fails to find a codeword jointly typical
with its input when the reference channel law is in effect.
\end{itemize}

A distortion violation event can also occur if there is no typicality
failure but the distortion is still too high.  Letting
\begin{equation} 
\edfail = \encSNotTyp \cup \encFail \cup \encChFail \cup \decFail
\label{eq:edfail}
\end{equation}
denote the typicality failure event, we have then that the probability
of a distortion violation can be expressed as
\begin{multline}
\Pr[\Edv] 
= \Pr[\Edv \mid \edfail]\cdot\Pr[\edfail]
+ \Pr[\Edv \mid \noedfail]\cdot\Pr[\noedfail] \\
\leq \Pr[\Edv \mid \noedfail] + \Pr[\edfail] \\
= \Pr\left[\Edv \mid \noedfail \right]
+ \Pr[\encSNotTyp]
+ \Pr[\encFail\mid\encSNotTyp^\comp]\\
+\Pr[\encChFail\mid\encSNotTyp^\comp,\encFail^\comp]
+\Pr[\decFail\mid\encSNotTyp^\comp,\encFail^\comp,\encChFail^\comp].
\label{eq:enc_error}
\end{multline}

First, according to well-known properties of typical sequences
\cite{cover}, by choosing $n$ large enough we can make
\begin{align} 
\Pr[\encSNotTyp] 
    &\leq \epsilon/4 \label{eq:prob_encsnottyp} \\
\Pr[\encChFail \mid \encSNotTyp^\comp,\encFail^\comp] 
    &\leq \epsilon/4. \label{eq:prob_encchfail} 
\end{align}

Second, provided that the source is typical, the probability that the
encoder fails to find a sequence $\ful{\naux}\in\admissC$ jointly
typical with the source follows from \eqref{eq:Rpdef} as
\begin{equation}
\Pr[\encFail\mid\encSNotTyp^\comp] \leq 2^{-n[R' - I(\nSrc;\nAux)]} =
2^{-n\gamma} 
\label{eq:degfail}
\end{equation}
from standard joint typicality arguments.  

Third,
\begin{equation} 
\Pr[\decFail \mid \encSNotTyp^\comp,\encFail^\comp,\encChFail^\comp]
\leq 2^{-n\gamma} + \epsilon/4.  
\label{eq:decfail}
\end{equation}
Indeed, using standard joint typicality results, the probability that
the sequence $\ful{\nChOut}$ presented to the decoder is not
$\delta$-strongly jointly typical with the correct codeword
$\ful{\nAux}$ selected by the encoder can be made smaller than
$\epsilon/4$ for $n$ large enough, and the probability of it being
strongly jointly typical with any other admissible codeword is, using
\eqref{eq:gammadef} with \eqref{eq:Rdef}, at most
\begin{equation*} 
2^{-n[I(\nAux;\nChOut)-\cbkR]} \le 2^{-n\gamma}.
\end{equation*}

Fourth,
\begin{equation}
\Pr\left[\Edv \mid \noedfail \right] = 0. 
\label{eq:prob_bad_system}
\end{equation}
Indeed, provided there are no typicality failures, the pair
$(\ful{\nSrc},\ful{\nChOut})$ must be strongly jointly typical, so by
the standard properties of strong joint typicality,
\begin{align*}
\frac{1}{n} \sum_{i=1}^n \diste(\nSrc_i,\nChIn_i) 
&\leq E[\diste(\nSrc,\nChIn)] + \delta \cdot \dmax{1}\\
\frac{1}{n} \sum_{i=1}^n \distr(\nSrc_i,\sedeci{i}{\nAux_i}) 
&\leq E[\distr(\nSrc,\sedec{\nAux})] + \delta \cdot \dmax{2},
\end{align*}
where $\dmax{1}$ and $\dmax{2}$ are bounds defined via
\begin{align}
\dmax{1} &= \sup_{(\nsrc,\nchin)\in\srcAlph\times\chinAlph}
 \diste(\nsrc,\nchin) \label{eq:dmax1-def}\\
\dmax{2} &= \sup_{(\nsrc,\nsrch)\in\srcAlph\times\srcAlph}
 \distr(\nsrc,\nsrch). \label{eq:dmax2-def}
\end{align}
Thus, choosing $\delta$ such that
\begin{equation*}
\delta < \max \left(\frac{\epsilon}{\dmax{1}},
\frac{\epsilon}{\dmax{2}} \right) 
\end{equation*}
and making $n$ large enough we obtain \eqref{eq:prob_bad_system}.

Finally, using \eqref{eq:prob_encsnottyp}, \eqref{eq:prob_encchfail},
\eqref{eq:degfail}, \eqref{eq:decfail}, and \eqref{eq:prob_bad_system}
in \eqref{eq:enc_error} we obtain
\begin{equation}
\label{eq:exdist_err}
\Pr[\Edv] \leq 3 \epsilon/4 + 2 \cdot 2^{-n\gamma}
\end{equation}
which can be made less than $\epsilon$ for $n$ large enough.
Thus $\Pr[\exdiste]\rightarrow0$ and, when the reference channel is in
effect, $\Pr[\exdistr]\rightarrow0$.

\noindent\qed

\subsection{Converse Part: Necessity}
\label{sec:converse_part}

Here we show that if there exists an
authentication system where the pair
$(\Diste,\Distr)$ is in the achievable distortion
region, then there exists a distribution $p(\naux|\nsrc)$ and
functions $\sedec{\cdot}$, $\senc{\cdot,\cdot}$ satisfying
\eqref{eq:thm}.  In order to apply previously developed tools, it is
convenient to define the rate-function 
\begin{multline}
\label{eq:def-rate-func}
R^*(\Diste,\Distr) \defeq\\
\sup_{\textnormal{\parbox{1.75in}{\begin{center}
$p(\nAux|\nSrc),f:\auxAlph\times\srcAlph\mapsto\chinAlph,g:\auxAlph\mapsto\srcAlph$\\
      \mbox{$: E[\diste(\nSrc,\senc{\nAux,\nSrc})]\leq\Diste,
      E[\distr(\nSrc,\sedec{\nAux})]\leq\Distr$}\end{center}}}}
      I(\nAux;\nChOut) - I(\nSrc;\nAux).
\end{multline}
Note that $R^*(\Diste,\Distr) \geq 0$ if and
only if the conditions in \eqref{eq:thm} are satisfied.  Thus our
strategy is to assume that the sequence of 
encoding and decoding functions discussed in \secref{sec:codethms}
exist with $\lim_{n\rightarrow\infty}\Pr[\undetErr]=0$,
$\lim_{n\rightarrow\infty}\Pr[\exdiste]=0$, and---when the reference
channel is in effect---$\lim_{n\rightarrow\infty}\Pr[\exdistr]=0$.
We then show that these functions imply that $R^*(\Diste,\Distr) \geq
0$ and hence \eqref{eq:thm} is satisfied.

To begin we note that it suffices to choose $\sedec{\cdot}$ to be the
minimum distortion estimator of $\nSrc$ given $\nAux$.  Next, 
by using techniques from \cite{gelfand_1980} or
by directly applying \cite[Lemma 2]{rjb_bc_gw_preprint} it is possible to
prove that allowing $\nChIn$ to be non-deterministic has no advantage,
i.e.,
\begin{multline}
\label{eq:det-good-enough}
R^*(\Diste,\Distr) \geq\\
\sup_{\textnormal{\parbox{1.75in}{\begin{center}
$p(\nAux|\nSrc),p(\nChIn|\nAux,\nSrc):$\\
      \mbox{$E[\diste(\nSrc,\nChIn)]\leq\Diste,  
      E[\distr(\nSrc,\sedec{\nAux})]\leq\Distr$}\end{center}}}}
      I(\nAux;\nChOut) - I(\nSrc;\nAux).
\end{multline}
Arguments similar to those in \cite{gelfand_1980} and
\cite[Lemma 1]{rjb_bc_gw_preprint} show that $R^*(\Diste,\Distr)$ is
monotonically non-decreasing and concave in $(\Diste,\Distr)$.  These
properties will later allow us to make use of the following lemma,
whose proof follows readily from that of Lemma~4 in \cite{gelfand_1980}:
\begin{lemma}
\label{gelfand_lemma}
For arbitrary random variables $V,A_1,A_2,\dots,A_n$ and a sequence of
i.i.d.\ random variables $\nSrc_1,\nSrc_2,\dots,\nSrc_n$,
\begin{multline}
\sum_{i=1}^n
\left[ I(V,A_1^{i-1},S_{i+1}^n;A_i)-I(V,A_1^{i-1},S_{i+1}^n;S_i)
\right] \\
\geq I(V;\ful{A}) - I(V;\ful{S}).
\end{multline}
\end{lemma}

As demonstrated by the following Lemma, a suitable $\nAux_i$ is
\begin{equation}
\nAux_i = (\ful{\nSrch},\nChOut_{1}^{i-1},\nSrc_{i+1}^n).
\label{eq:def_nauxdeg}
\end{equation}
\begin{lemma}
\label{lem:umarkov}
The choice of $\nAux_i$ in \eqref{eq:def_nauxdeg} satisfies the 
Markov relationship
\begin{equation} 
\nChOut_i \leftrightarrow (\nSrc_i, \nChIn_i) \leftrightarrow \nAux_i.
\label{eq:umarkov}
\end{equation}
\end{lemma}
\begin{proof}
It suffices to note that
\begin{align} 
p(\nchout_i|\nchin_i,\nsrc_i) 
&= p(\nchout_i|\nchin_i) 
= \frac{p(\nchout_1^i|\ful{\nchin})}{p(\nchout_1^{i-1}|\ful{\nchin})} 
= \frac{p(\nchout_1^i|\ful{\nchin},\ful{\nsrc})}
       {p(\nchout_1^{i-1}| \ful{\nchin},\ful{\nsrc})}
  \label{eq:from_memless_ch:a} \\  
&= \frac{p(\nchout_1^i|\ful{\nchin},\ful{\nsrch} ,\ful{\nsrc})}
       {p(\nchout_1^{i-1}|\ful{\nchin},\ful{\nsrch},\ful{\nsrc})}
= p(\nchout_i|\ful{\nchin},\ful{\nsrc},\ful{\nsrch},\nchout_1^{i-1})
\label{eq:from_other_m_cond}
\end{align}
where the equalities in \eqref{eq:from_memless_ch:a} follow from the
memoryless channel model, and the first equality in
\eqref{eq:from_other_m_cond} follows from the fact that the system
generates authentic reconstructions so \eqref{eq:estmarkov} holds.
Thus, \eqref{eq:from_other_m_cond} implies the Markov relationship
\begin{equation}
\nChOut_i \leftrightarrow (\nChIn_i,\nSrc_i) \leftrightarrow
(\nChIn_1^i,\nChIn_{i+1}^n,\nSrc_1^i,\nSrc_{i+1}^n,\nChOut_1^{i-1},\ful{\nSrch}),  
\label{eq:markov:almost}
\end{equation}
which by deleting selected terms from the right hand side yields
\eqref{eq:umarkov}.
\end{proof}

Next, we combine these results to prove the converse part of
\thrmref{th:main} except for the cardinality bound on $\auxAlph$ which
is derived immediately thereafter.
\begin{lemma}
\label{lem:deg_prod_space}
If a sequence of encoding and decoding
functions $\encoder(\cdot)$ and 
$\xdecn{\cdot}$ exist such that the decoder can generate authentic
reconstructions achieving the distortion pair $(\Diste,\Distr)$ when the
reference channel is in effect then
\begin{equation}
R^*(\Diste,\Distr) \geq 0.
\label{eq:deg_prod_space}
\end{equation}
\end{lemma}
\begin{proof}
Define $\Distei$ and $\Distri$ as the component-wise distortions
between $\nSrc_i$ and $\nChIn_i$ and between $\nSrc_i$ and
$\nSrch_i$.  We have the following chain of inequalities:
\begin{align}
R^*(\Diste,\Distr) &= R^*\left(\frac{1}{n}\sum_{i=1}^n \Distei,
\frac{1}{n}\sum_{i=1}^n \Distri\right)\\
\label{eq:rstar-conc}
&\geq \frac{1}{n} \sum_{i=1}^n R^*(\Distei,\Distri)\\
\label{eq:rstar-bigger}
&\geq \frac{1}{n}
\sum_{i=1}^n[I(\nAux_i;\nChOut_i)-I(\nAux_i;\nSrc_i)] \\
\label{eq:use-gf-lemma}
&\geq \frac{1}{n} \left[
I(\ful{\nSrch};\ful{\nChOut})-I(\ful{\nSrch};\ful{\nSrc}) \right]\\
&=\frac{1}{n} \left[
H(\ful{\nSrch}|\ful{\nSrc})-H(\ful{\nSrch}|\ful{\nChOut}) \right] \\
&\geq -\frac{1}{n} H(\ful{\nSrch}|\ful{\nChOut})\\
\label{eq:fin-apply-fano}
&\geq -\frac{1}{n} - \Pr[\xdecn{\ful{\nChOut}}\neq\ful{\nSrch}]
\log\card{\srcAlph}.
\end{align}

The concavity of $R^*(\Diste,\Distr)$ yields \eqref{eq:rstar-conc}.
To obtain \eqref{eq:rstar-bigger}, we combine \lemref{lem:umarkov}
with \eqref{eq:det-good-enough}.  Next, to obtain
\eqref{eq:use-gf-lemma}, let $V = \ful{\nSrch}$ and $A_i=\nChOut_i$ to
apply \lemref{gelfand_lemma} with $\nAux_i$ chosen according to
\eqref{eq:def_nauxdeg}.  Fano's inequality yields
\eqref{eq:fin-apply-fano}.

Finally, using (in order) Bayes' law,
\eqref{eq:undetErr-def}, and \eqref{eq:d2def}, we obtain
\begin{align}
\Pr[\xdecn{\ful{\nChOut}}&\neq\ful{\nSrch}] = \Pr[\undetErr] \notag\\
& \hspace{-40pt} + \Pr[\{\xdecn{\ful{\nChOut}}\neq\ful{\nSrch}\} \cap
\{\xdecn{\ful{\nChOut}}=\dfail\}]\\
&\leq \Pr[\undetErr] +
\Pr[\{\xdecn{\ful{\nChOut}}=\dfail\}]\\ 
&\leq \Pr[\undetErr] + \Pr[\exdistr].
\label{eq:key_fano_term}
\end{align}
Therefore exploiting that the system generates an authentic
reconstruction ($\lim_{n\rightarrow\infty}\Pr[\undetErr] = 0$) of the
right distortion ($\lim_{n\rightarrow\infty}\Pr[\exdistr] = 0$) and
that the alphabet of $\nSrc$ is finite, we have that
\eqref{eq:fin-apply-fano} and \eqref{eq:key_fano_term} imply
\eqref{eq:deg_prod_space}.
\end{proof}

The following proposition bounds the cardinality of $\auxAlph$.
\begin{prop}
\label{prop:card}
Any point in the achievable distortion region defined by
\eqref{eq:thm} can be attained with $\nAuxDeg$ distributed over an
alphabet $\auxAlph$ of cardinality at most $(\card{\srcAlph} +
\card{\chinAlph} + 3)\cdot\card{\srcAlph}\cdot\card{\chinAlph}$ with
$p(\nchin|\naux,\nsrc)$ singular or over an
alphabet $\auxAlph$ of cardinality at most $\card{\srcAlph} +
\card{\chinAlph} + 3$ if $p(\nchin|\naux,\nsrc)$ is not required to be
singular. 
\end{prop}

\begin{proof}
This can be proved using standard tools from convex set theory.
Essentially, we define a convex set of continuous functions
$f_j({\mathbf p})$ where ${\mathbf p}$ represents a distribution of
the form $\Pr(\nSrc=\nsrc,\nChIn=\nchin|\nAux=\naux)$ and the
$f_j(\cdot)$ functions capture the features of the distributions
relevant to \eqref{eq:thm}.  According to Carath\'{e}odory's Theorem
\cite[Theorem 14.3.4]{cover}, \cite{it:wyner_1975}, there exist
$j_{\max}$ +1 distributions ${\mathbf p}_1$ through ${\mathbf
p}_{\textnormal{$j_{\max}$ +1}}$ such that any vector of function values,
$(f_1({\mathbf p'}), f_2({\mathbf p'}), \dots,
f_{\textnormal{$j_{\max}$}}({\mathbf p'}))$, achieved by some
distribution ${\mathbf p'}$ can be achieved with a convex combination of
the ${\mathbf p}_i$ distributions.  Since each distribution
corresponds to a particular choice for $\nAux$, at most $j_{\max}$ + 1
possible values are required for $\nAux$.  Specifically, the desired
cardinality bound for our problem can be proved by making the
following syntactical modifications to the argument in \cite[bottom
left of p.~634]{it:ahlswede_1976}:

\begin{enumerate}

\item Replace $\Pr(X = x \mid U = u)$ with $\Pr(\nSrc = \nsrc,\nChIn =
\nchin \mid \nAux = \naux)$ which is represented by the notation
$\mathbf{p}$.

\item Choose 
\begin{equation}
f_j(\mathbf{p}) = 
\sum_{\nchin} \Pr(\nSrc = j,\nChIn = \nchin \mid \nAux = \naux)
\end{equation}
for $j \in \{1,2,\dots,n\}$ where $n = \card{\srcAlph}$. 

\item Choose 
\begin{multline} 
f_{n+1}(\mathbf{p}) =\\ \sum_{\nsrc} \sum_{\nchin} 
\diste(\nchin,\nsrc) \,
\Pr(\nSrc = \nsrc,\nChIn = \nchin \mid \nAux = \naux).
\end{multline}

\item Choose 
\begin{multline}
f_{n+2}(\mathbf{p}) =\\ \sum_{\nsrc} \sum_{\nchin}
\distr(\sedec{\naux},\nsrc)\,
\Pr(\nSrc = \nsrc, \nChIn = \nchin \mid \nAux = \naux).
\end{multline}

\item Choose 
\begin{multline} 
f_{n+3}(\mathbf{p}) = \sum_{\nsrc} \left[\sum_{\nchin}
 \Pr(\nSrc = \nsrc,\nChIn = \nchin \mid \nAux = \naux) 
  \cdot\right.\\ 
\left.\ \ \log \left(\sum_{\nchin}
\Pr(\nSrc = \nsrc,\nChIn = \nchin \mid \nAux = \naux) \right)\right].
\end{multline}

\item Let 
\begin{multline*}
m(s,u,x,y) = \\\Pr(\nChOut=\nchout \mid \nChIn=\nchin)
\Pr(\nSrc = \nsrc,\nChIn = \nchin \mid \nAux = \naux)
\end{multline*}
and choose
\begin{multline}
f_{n+4}(\mathbf{p}) = \sum_{\nchout} \left[
\left(\sum_{\nchin}\sum_{\nsrc} m(s,u,x,y)\right) \right.\cdot\\
\left. \left(\sum_{\nchin}\sum_{\nsrc} \log m(s,u,x,y) \right)\right].
\end{multline}

\item Choose
\begin{equation}
f_{n+5+j}(\mathbf{p}) = \sum_{\nsrc} \Pr(\nSrc = \nsrc,\nChIn =
j \mid \nAux = \naux)
\end{equation}
for $j \in \{1,2, \dots,\card{\chinAlph}\}$.

\end{enumerate}

Since the $f_j(\mathbf{p})$ determine $\Pr[\nSrc = \nsrc]$ (and
therefore $H(\nSrc)$ as well), $\Diste$, $\Distr$,
$H(\nSrc|\nAuxDeg)$, $H(\nChOut|\nAuxDeg)$, and $\Pr[\nChIn = \nchin]$
(and therefore $\Pr[\nChOut=\nchout]$ and $H(\nChOut)$ also), they can
be used to identify all points in the distortion region.  According to
\cite[Lemma 3]{it:ahlswede_1976}, for every point in this region
obtained over the alphabet $\auxAlph$ there exists a $U^*$ from
alphabet $\auxAlph^*$ with cardinality $\card{\auxAlph^*}$ at most one
greater than the dimension of the space spanned by the vectors $f_i$.
The $f_i$ corresponding to $\Pr[\nSrc=\nsrc]$ and $\Pr[\nChIn=\nchin]$
contribute $\card{\srcAlph}-1$ and $\card{\chinAlph}-1$ dimensions
while the other $f_i$ contribute four more dimensions.  Thus it
suffices to choose $\card{\auxAlph^*} \leq \card{\chinAlph} +
\card{\srcAlph} + 3$.  Note that this cardinality bound applies to the
general case where $\nChIn$ is not necessarily a deterministic
function of $\nSrc$ and $\nAux^*$.

By directly applying \cite[Lemma 2]{rjb_bc_gw_preprint} to each pair
$(\naux^*,\nsrc)$ in $\auxAlph^*\times\srcAlph$, we can split each $\naux^*$
into $\card{\chinAlph}$ new symbols $\naux^{**}$ such that the mapping
from $(\naux^{**},\nsrc)$ to $\nchin$ is deterministic.  The new
auxiliary random variable $\nAux^{**}$ takes values over the alphabet
$\auxAlph^{**}$ where
\begin{equation}
\card{\auxAlph^{**}} = \card{\auxAlph^*}\cdot\card{\srcAlph}\cdot\card{\chinAlph}
= (\card{\chinAlph} + \card{\srcAlph} +
3)\cdot\card{\srcAlph}\cdot\card{\chinAlph}.
\end{equation}
Furthermore, this process does not change the distortion or violate
the mutual information constraint.  Thus a deterministic mapping from
the source and auxiliary random variable to the channel input can be
found with no loss of optimality provided a potentially larger alphabet is
allowed for the auxiliary random variable.
\end{proof}

We next apply Theorem~\ref{th:main} to two example scenarios of
interest---one discrete and one continuous.

\section{Example: the Binary-Hamming Scenario}
\label{sec:binary_hamming}

In some applications of authentication, the content of interest is
inherently discrete.  For example, we might be interested in
authenticating a passage of text, some of whose characters may have
been altered in a benign manner through errors in optical character
recognition process or error-prone human transcription during
scanning.  Or the alterations might be by the hand of human editor
whose job it is to correct, refine, or otherwise enhance the
exposition in preparation for its publication in a paper, journal,
magazine, or book.  Or the alternations may be the result of an
attacker deliberately tampering with the text for the purpose of
distorting its meaning and affecting how it will be interpreted.

As perhaps the simplest model representative of such discrete
problems, we now consider a symmetric binary source with a binary
symmetric reference channel.  Specifically, we model the source as an
i.i.d.\ sequence where each $\nSrc_i$ is a Bernoulli($1/2$) random
variable\footnote{We adopt the convention that all Bernoulli random
variables take values in the set $\{0,1\}$.} and the reference channel
output is $\nChOut_i = \nChIn_i \oplus N_i$, where $\oplus$ denotes
modulo-$2$ addition and where $\ful{N}$ is an i.i.d.\ sequence of
Bernoulli($\crossProb$) random variables.  Finally, we adopt the
Hamming distortion measure:
\begin{equation*}
d(a,b) = 
\begin{cases} 
0, & \text{ if $a = b$}\\
1, & \text{ otherwise }.
\end{cases}
\end{equation*}

For this problem, a suitable auxiliary random variable is
\begin{equation}
\nAuxDeg = \left\{ \nSrc \oplus (A \cdot \genrv) \oplus 
[(1-A) \cdot V]\right\} + 2 \cdot (1-A),
\label{eq:bin_dist_def:u}
\end{equation}
where $A$, $\genrv$, and $V$ are Bernoulli $\alpha$, $\tau$, and $\nu$
random variables, respectively, and are independent of each other and
$\nSrc$ and $N$.  Without loss of generality, the parameters 
$\tau$ and $\nu$ are restricted to the range $(0,1/2)$.  Note that
$\auxAlph=\{0, 1, 2, 3\}$.

The encoder function $\nChIn=f(\nSrc,\nAux)$ is, in turn, given by
\begin{equation}
\nChIn = \begin{cases} 
         \nAuxDeg, & \text{if $\nAuxDeg \in \{0, 1\}$} \\
         \nSrc, & \text{if $\nAuxDeg \in \{2, 3\}$},
	 \end{cases}
\label{eq:bin_dist_def:x}
\end{equation}
from which it is straightforward to verify via
\eqref{eq:bin_dist_def:u} that the encoding distortion is
\begin{equation} 
\Diste = \alpha\tau.
\label{eq:Diste-bh}
\end{equation}

The corresponding decoder function $\nSrch=g(\nAux)$ takes the
form
\begin{equation} 
\nSrch = \nAux \bmod 2,
\end{equation}
from which it is straightforward to verify via
\eqref{eq:bin_dist_def:u} that the reconstruction distortion is
\begin{equation} 
\Distr = \alpha\tau + (1-\alpha)\nu.
\label{eq:Distr-bh}
\end{equation}

\begin{figure}[tbp]
\centering
\psfrag{TITLE}{}
\psfrag{D1}{\LARGE$\Diste$}
\psfrag{D2}{\LARGE$\Distr$}
\includegraphics[angle=0,width=3.0in]{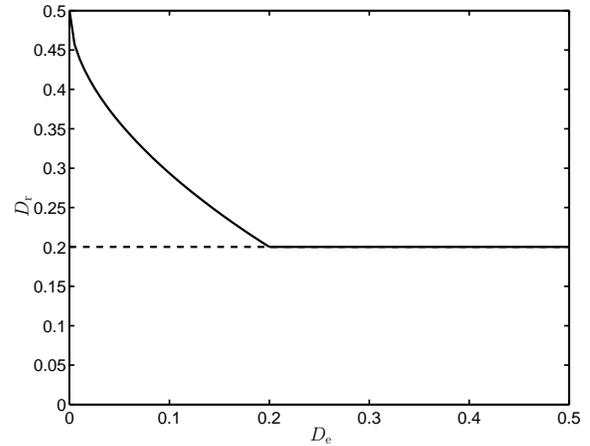}
\caption{The solid curve represents the frontier of the achievable
distortion region for a binary symmetric source and a binary symmetric
reference channel with cross-over probability $\crossProb=0.2$.  This
plot reflects the system behavior when the reference channel is in
effect.  The dashed line represents the boundary of the larger
distortion region achievable when authentication is not required.
\label{fig:ham_bin_reg}}
\end{figure}

In addition, $I(\nAuxDeg;\nSrc)$ takes the form
\begin{align}
I(\nAuxDeg;\nSrc) &= H(\nSrc) - H(\nSrc|\nAuxDeg)\notag\\
&= H(\nSrc) - H(\nSrc,A|\nAuxDeg) + H(A|\nAuxDeg,\nSrc)\notag\\
&= H(\nSrc) - H(\nSrc|\nAuxDeg,A) - H(A|\nAuxDeg) + H(A|\nAuxDeg,\nSrc)\notag\\
&= 1 - \alpha\cdot h(\tau) - (1-\alpha)\cdot h(\nu),
\label{eq:ius-val}
\end{align}
where the second and third equalities follow from the entropy chain
rule, where the last two terms on the third line are zero
because knowing $\nAuxDeg$ determines $A$, and where the last equality
follows from \eqref{eq:bin_dist_def:u}, with $h(\cdot)$ denoting the
binary entropy function, i.e., $h(q)=-q\log q - (1-q)\log(1-q)$ for
$0\le q\le 1$.  Similarly, $I(\nAux;\nChOut)$ takes the form
\begin{align}
I(\nAuxDeg;\nChOut) &= H(\nChOut) - H(\nChOut|\nAuxDeg)\notag\\
&= H(\nChOut) - H(\nChOut,A|\nAuxDeg) + H(A|\nAuxDeg,\nChOut)\notag\\
&= H(\nChOut) - H(\nChOut|\nAuxDeg,A) - H(A|\nAuxDeg) + H(A|\nAuxDeg,\nChOut)\\
&= 1 - \alpha\, h(\crossProb) - (1-\alpha) 
   h\left(\crossProb(1- \nu) + (1-\crossProb)\nu\right).
\label{eq:iuy-val}
\end{align}
For a fixed $\crossProb$, varying the parameters $\alpha$, $\nu$,
and $\tau$ such that \eqref{eq:iuy-val} is at least as big as
\eqref{eq:ius-val} as required by \eqref{eq:thm:a} generates the
achievable distortion region shown in Fig.~\ref{fig:ham_bin_reg}.
Note from \eqref{eq:iuy-val}, \eqref{eq:ius-val}, \eqref{eq:Diste-bh}
and \eqref{eq:Distr-bh} that the boundary point $\Diste = \Distr =
\crossProb$, in particular, is obtained by the parameter values
$\alpha=1$ and $\tau=\crossProb$ (with any choice of $\nu$).
Numerical optimization over all $p(\nauxdeg|\nsrc)$ and all (not
necessarily singular) $p(\nchin|\nsrc,\naux)$ with the alphabet size
$\card{\auxAlph} = 
7$ chosen in accordance with Proposition~\ref{prop:card} confirms that
Fig.~\ref{fig:ham_bin_reg} captures all achievable distortion pairs.

For comparison, we can also develop the achievable distortion region
when authentication is not required.  In this setting the goal is to
provide a representation of the source which allows a decoder to
obtain a good reconstruction from the reference channel output while
keeping the encoding distortion small.  Although in general hybrid
analog-digital coding schemes can be used \cite{rjb_bc_gw_preprint},
optimality can also be achieved without any coding in the
binary-Hamming case and thus all points in the region $\Diste\geq0$
and $\Distr\geq\crossProb$ are achievable, as also shown in
Fig.~\ref{fig:ham_bin_reg}.  Thus we see that the requirement that
reconstructions be authentic strictly decreases the achievable
distortion region as shown in Fig~\ref{fig:ham_bin_reg}.

\section{Example: the Gaussian-Quadratic Scenario}
\label{sec:gaussian}

In some other applications of authentication, the content of interest
is inherently continuous.  Examples involve sources such as imagery,
video, or audio.  In addition to tampering attacks, such content may
encounter degradations as a result of routine handling that includes
compression, transcoding, resampling, printing, and scanning, as well
as perturbations from editing to enhance the content.

As perhaps the simplest model representative of such continuous
problems, we consider a white Gaussian source with a white Gaussian
reference channel.  Specifically, we model the source as an i.i.d.\
Gaussian sequence where each $\nSrc_i$ has mean zero and variance
$\sigma_{\nSrc}^2$, and the independent reference channel noise as an
i.i.d.\ sequence whose $i$\/th element $N_i$ has mean zero and
variance $\sigma_N^2$.  Furthermore, we adopt the quadratic distortion
measure $d(a,b) = (a-b)^2$.

While our proofs in Section~\ref{sec:proofs} exploited that our
signals were drawn from finite alphabets and that all distortion
measures were bounded to simplify our development, the results can be
generalized to continuous-alphabet sources with unbounded distortion
measures using standard methods.  In the sequel, we assume without
proof that the coding theorems hold for Gaussian sources with
quadratic distortion.  Since it appears difficult to obtain a
closed-form expression for the optimal distribution for
$\nAuxDeg$,\footnote{An analysis using calculus of variations
suggests that the optimal distribution is not even Gaussian.} we
instead develop good inner and outer bounds on the boundary of the
achievable distortion region.

\subsection{Unachievable Distortions: Inner Bounds}
\label{sec:gaussian:unachievable}

To derive an inner bound, we ignore the requirement that
reconstructions be authentic, i.e., satisfy \eqref{eq:estmarkov}, and
study the distortions possible in this case.  

For a given constraint on the power $P$ input to the reference
channel, it is well-known that the minimum possible source
reconstruction distortion $\Distr$ achievable from the output of the
channel can be achieved without either source or channel coding in
this Gaussian scenario, and the resulting distortion is
\begin{equation}
\Distr = \frac{\sigma_N^2 \sigma_{\nSrc}^2}{\sigma_N^2 + P}.
\label{eq:d2:tx}
\end{equation}
Moreover, for a scheme with encoding distortion $\Diste$, the
Cauchy-Schwarz inequality implies that $P$ is
bounded according to
\begin{multline} 
P = E[\nChIn^2] = E[(\nChIn-\nSrc + \nSrc)^2]
= E[(\nChIn-\nSrc)^2] + E[\nSrc^2]\\
 + 2E[(\nChIn-\nSrc)\nSrc] 
\leq \Diste + \sigma_{\nSrc}^2 + 2\sqrt{\Diste\sigma_{\nSrc}^2},
\label{eq:max_nchout_var} 
\end{multline}
where equality holds if and only if $\nChIn = \left(1 +
\sqrt{\Diste/\sigma_{\nSrc}^2}\right)\nSrc$.  Thus, substituting
\eqref{eq:max_nchout_var} into \eqref{eq:d2:tx} yields the inner bound
\begin{equation}
\Distr = \frac{\sigma_N^2 \sigma_{\nSrc}^2}{\sigma_N^2 +
\left(\sqrt{\Diste} + \sigma_{\nSrc}\right)^2}.
\label{eq:d2:lb}
\end{equation}

\subsection{Achievable Distortions: Outer Bounds}
\label{sec:gaussian:ach_dist}

To derive outer bounds we will consider codebooks where
$(\nSrc,\nAux,\nChIn)$ are jointly Gaussian.  Since it is sufficient
to consider $\nChIn$ to be a deterministic function of $\nAux$ and
$\nSrc$, the innovations form
\begin{subequations}
\label{eq:gauss_innov_form}
\begin{align}
\genrv &\sim N(0,\sigma_{\genrv}^2), \ \ E[\genrv\nSrc] = 0\\
\nAux &= a \nSrc + c \genrv\\
\nChIn &= b \nAux + d \genrv
\end{align}
\end{subequations}
conveniently captures the desired relationships.\footnote{It can be
shown that choosing either $a=1$ or $c=1$ incurs no loss of
generality.}  We examine two regimes: a low $\Diste$ regime in which
we restrict our attention to the parameterization $(a,b,c,d) =
(1,1,1/\alpha,1)$, and a high $\Diste$ regime in which we restrict our
attention to the parameterization $(a,b,c,d) = (1,\beta,1,0)$.  As
we will see, time-sharing between these parameterizations yields
almost the entire achievable distortion region for Gaussian codebooks.

\subsubsection*{Low $\Diste$ Regime}

We obtain an encoding that is asymptotically good at low
$\Diste$ by using a distribution with structure similar to that used
to achieve capacity in the related problem of information embedding
\cite{costa_83}.  In the language of \cite{chen_wornell_2001}, the
encoding process involves distortion-compensation.  In particular, the
source is amplified by a factor $1/\alpha$, quantized to the nearest
codeword, attenuated by $\alpha$, and then a fraction of the resulting
quantization error is added back to produce the final encoding, i.e.,
\begin{equation} 
\ful{\nChIn} = \alpha Q[\ful{\nSrc}/\alpha] + (1-\alpha) ( \ful{\nSrc}
- \alpha Q[\ful{\nSrc}/\alpha])
\end{equation}
where $Q[\cdot]$ denotes the quantizer function.

With this encoding structure, it is convenient to make the assignment
$\ful{\nAux} = \alpha Q[\ful{\nSrc}/\alpha]$, so that we may write
\begin{align} 
\nAux &= \nSrc + \genrv/\alpha \label{eq:dc-aux}\\
\nChIn &= \nAux + (1-\alpha)(\nSrc-\nAux) =  \nSrc + \genrv \label{eq:dc-chin}
\end{align}
where $\genrv$ is a Gaussian random variable with mean zero and
variance $\sigma_{\genrv}^2$ independent of both the source $\nSrc$
and the reference channel noise $N$.

We choose $\sedec{\cdot}$ to be the minimum mean-square estimate of
$\nSrc$ given $\nAux$.  Thus the resulting distortions are, via
\eqref{eq:dc-aux} and \eqref{eq:dc-chin},
\begin{equation} 
\Diste = E[(\nChIn-\nSrc)^2] 
       = E[(\nSrc + \genrv - \nSrc)^2] 
       = \sigma_{\genrv}^2 
\label{eq:d1:bnd} 
\end{equation}
and, in turn,
\begin{align}
\Distr &= E[\nSrc^2]\left(1 -
\frac{E[\nSrc\nAux]^2}{E[\nSrc^2]E[\nAux^2]}\right)\notag\\
&= \frac{\sigma_{\nSrc}^2
   (\sigma_{\genrv}^2+\alpha^2\sigma_{\nSrc}^2 ) -
   \alpha^2\sigma_{\nSrc}^4}{\sigma_{\genrv}^2+\alpha^2\sigma_{\nSrc}^2
   }\notag\\
&= \frac{\sigma_{\nSrc}^2
   \Diste}{\Diste+\alpha^2\sigma_{\nSrc}^2}.   
\label{eq:d2:bnd}
\end{align}

To show that distortions \eqref{eq:d1:bnd} and \eqref{eq:d2:bnd} are
achievable requires proving that \eqref{eq:thm:a} holds.  In
\cite{costa_83}, the associated difference of mutual informations is
computed (using slightly different notation) as
\begin{multline}
I(\nAux;\nChOut)-I(\nSrc;\nAux) =\\
\frac{1}{2}\log\frac{\sigma_{\genrv}^2(\sigma_{\genrv}^2 +
\sigma_{\nSrc}^2 +
\sigma_N^2)}{\sigma_{\genrv}^2\sigma_{\nSrc}^2(1-\alpha)^2 +
\sigma_N^2(\sigma_{\genrv}^2 + \alpha^2\sigma_{\nSrc}^2)}
\end{multline}
which implies that to keep the difference of mutual informations
nonnegative we need 
\begin{equation}
\sigma_{\genrv}^2(\sigma_{\genrv}^2 +
\sigma_{\nSrc}^2 +
\sigma_N^2) \geq \sigma_{\genrv}^2\sigma_{\nSrc}^2(1-\alpha)^2 +
\sigma_N^2(\sigma_{\genrv}^2 + \alpha^2\sigma_{\nSrc}^2).
\end{equation}
Collecting terms in powers of $\alpha$ yields
\begin{equation} 
\alpha^2(\sigma_{\genrv}^2\sigma_{\nSrc}^2 +
\sigma_{N}^2\sigma_{\nSrc}^2) - 2\alpha
\sigma_{\genrv}^2\sigma_{\nSrc}^2 - \sigma_{\genrv}^4 
 = (\alpha-r_+)(\alpha-r_-) \le 0
\label{eq:r12-poly}
\end{equation}
where 
\begin{align} 
r_+ &= \frac{1 +
\sqrt{1 + \sigma_{\genrv}^2/\sigma_{\nSrc}^2 +
\sigma_{N}^2/\sigma_{\nSrc}^2}}{1 +
\sigma_{N}^2/\sigma_{\genrv}^2} \ge 0 \label{eq:rp-def}\\
r_- &= \frac{1 -
\sqrt{1 + \sigma_{\genrv}^2/\sigma_{\nSrc}^2 +
\sigma_{N}^2/\sigma_{\nSrc}^2}}{1 +
\sigma_{N}^2/\sigma_{\genrv}^2} \le 0. \label{eq:rm-def}
\end{align}
Therefore to satisfy the mutual information constraint we need 
$r_- \leq \alpha \leq r_+$.

To minimize the distortions, \eqref{eq:d2:bnd} and \eqref{eq:d1:bnd}
imply we want $|\alpha|$ as large as possible subject to the
constraint \eqref{eq:r12-poly}.  Thus we choose $\alpha = r_+$, from which
we see that 
\begin{equation} 
\frac{\alpha_{\mathrm{auth}}}{\alpha_{\mathrm{ie}}} 
=\left(1+\sqrt{1+\frac{\sigma_{\genrv}^2+\sigma_N^2}{\sigma_{\nSrc}^2}}\right),
\end{equation}
where
$\alpha_{\mathrm{ie}}=\sigma_{\genrv}^2/(\sigma_{\genrv}^2+\sigma_N^2)$
is the corresponding information embedding scaling parameter
determined by Costa \cite{costa_83}.  Evidently, the scaling parameter
for the authentication problem is at least twice the scaling for
information embedding and
significantly larger when either the 
SNR $\sigma_{\nSrc}^2/\sigma_N^2$  or signal-to-(encoding)-distortion
ratio (SDR) $\sigma_{\nSrc}^2/\sigma_{\genrv}^2$ is small.

\subsubsection*{High $\Diste$ Regime}

An encoder that essentially amplifies the quantization of the
source to overcome the reference channel noise is asymptotically good
at high $\Diste$.  A system with this structure corresponds to
choosing the encoder random variables according to
\begin{align}
\nAuxDeg &= \nSrc + \genrv\\
\nChIn &= \beta \nAuxDeg.
\end{align}
In turn, choosing as $\sedec{\cdot}$ the minimum mean-square
error estimator of $\nSrc$ given $\nAuxDeg$ yields the distortions
\begin{align}
\Diste &= (1-\beta)^2\sigma_{\nSrc}^2 + \beta^2\sigma_{\genrv}^2
\label{eq:d1_high_d1}\\
\Distr &= 
\frac{\sigma_{\nSrc}^2\sigma_{\genrv}^2}{\sigma_{\nSrc}^2 + \sigma_{\genrv}^2}.
\label{eq:d2_high_d1} 
\end{align}

It remains only to determine $\beta$.  Since
\begin{equation} 
I(\nAuxDeg;\nSrc) =
\frac{1}{2}\log\frac{\sigma_{\nSrc}^2+\sigma_{\genrv}^2}{\sigma_{\genrv}^2}
\end{equation}
and
\begin{equation} 
I(\nAuxDeg;\nChOut) = 
\frac{1}{2}\log\frac{\beta^2 (\sigma_{\nSrc}^2 + \sigma_{\genrv}^2) +
\sigma_N^2}{\sigma_N^2}, 
\end{equation}
the mutual information constraint \eqref{eq:thm:a} implies that 
\begin{equation} 
\beta \geq
\sqrt{\frac{\sigma_{\nSrc}^2\sigma_N^2}{\sigma_{\genrv}^2(\sigma_{\nSrc}^2
+ \sigma_{\genrv}^2)}}.
\label{eq:beta_def}
\end{equation}

\subsection{Comparing and Interpreting the Bounds}

\begin{figure*}[tbp]
\centering
\psfrag{D1AX}{\LARGE\hspace{-30pt}\raisebox{-.05in}{$\Diste/\sigma_N^2$ (in dB)}}
\psfrag{D2AX}{\LARGE\hspace{-30pt}\raisebox{.1in}{$\Distr/\sigma_N^2$ (in dB)}}
\includegraphics[angle=0,width=5.5in]{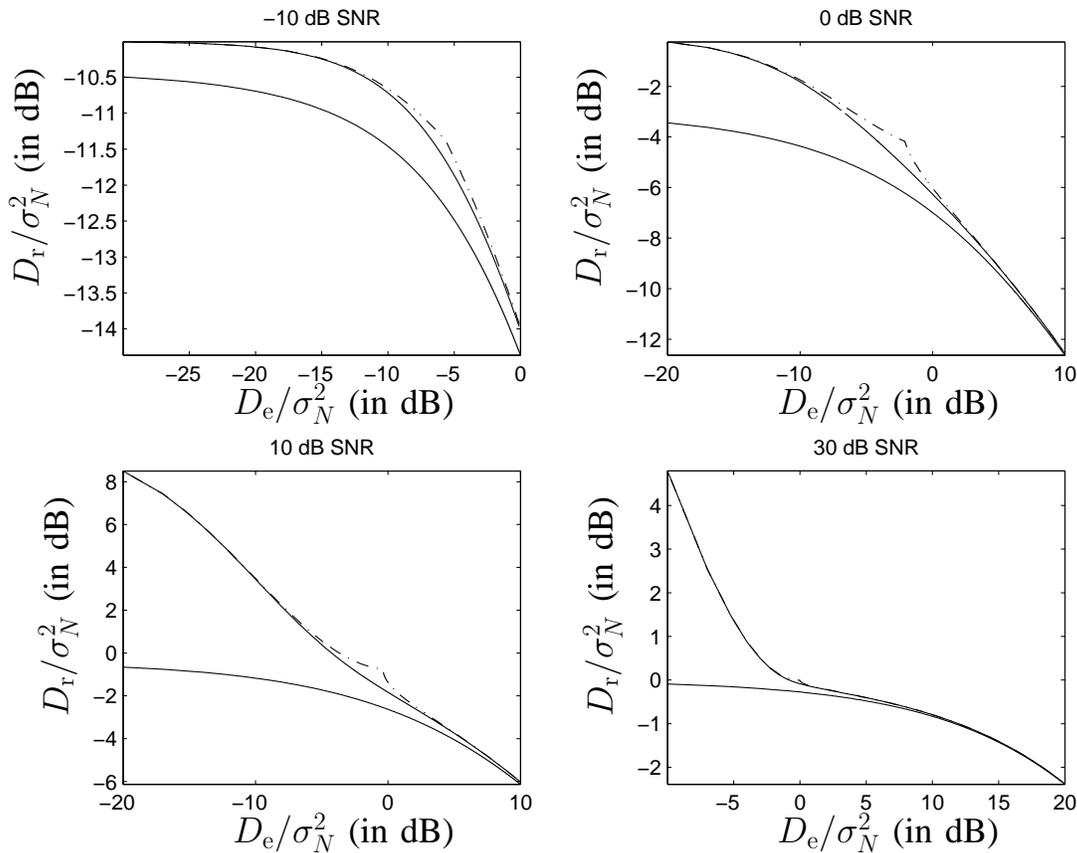}
\caption{Bounds on the achievable distortion region for the
Gaussian-quadratic problem.  The lowest solid curve is the inner bound
corresponding to the boundary of the achievable region when
reconstructions need not be authentic.  The numerically obtained upper
solid curve is the outer bound resulting from the use of Gaussian
codebooks.  The dashed curve corresponds to the lower convex envelope
of the simple low and high $\Diste$ analytic outer bounds derived in
the text. \label{fig:str_reg}}
\end{figure*}

Using \eqref{eq:d2:bnd} with $\alpha$ given by \eqref{eq:rp-def} and
varying $\sigma_{\genrv}^2$ yields one outer bound.  Using
\eqref{eq:d1_high_d1} and \eqref{eq:d2_high_d1} with
\eqref{eq:beta_def} and again varying $\sigma_{\genrv}^2$ yields the
other outer bound.  The lower convex envelope of this pair of outer
bounds is depicted in Fig.~\ref{fig:str_reg} at different SNRs.  To
see that the first and second outer bounds are asymptotically the best
achievable for low and high $\Diste$, respectively, we superimpose on
these figures the best Gaussian codebook performance, as obtained by
numerically optimizing the parameters in \eqref{eq:gauss_innov_form}.

By using \eqref{eq:d2:lb},
\eqref{eq:d2:bnd}, and \eqref{eq:d2_high_d1}, it is possible to show
that for any fixed $\Diste \ge \sigma_N^2$ the inner and outer bounds
converge asymptotically in SNR in the sense that
\[
\lim_{\textnormal{SNR}\rightarrow\infty}
\frac{D_{r,\textnormal{outer}}}{D_{r,\textnormal{inner}}} = 1
\]
where $D_{r,\textnormal{inner}}$ and $D_{r,\textnormal{outer}}$
represent the inner and outer bounds corresponding to the fixed value
of $\Diste$.  Thus, in this high SNR regime, Gaussian codebooks are
optimal, and \eqref{eq:d2:lb} accurately characterizes their
performance as reflected in Fig.~\ref{fig:str_reg}.

The figure also indicates (and it is possible to prove) that for any
fixed SNR, the inner and outer bounds converge asymptotically in
$\Diste$ in the sense that
\[
\lim_{\Diste\rightarrow\infty}
\frac{D_{r,\textnormal{outer}}(\Diste)}{D_{r,\textnormal{inner}}(\Diste)} = 1
\]
where $D_{r,\textnormal{inner}}(\Diste)$ and
$D_{r,\textnormal{outer}}(\Diste)$ represent the inner and outer
bounds as a function of the encoding distortion $\Diste$.  Evidently
in this high encoding distortion regime, $\Distr/\sigma_N^2$ can be
made arbitrarily small by using Gaussian codebooks and making
$\Diste/\sigma_N^2$ sufficiently large.  While this implies that, in
principle, there is no fundamental limit to how small we can make
$\Distr$ by increasing $\Diste$ through amplification of the source,
in practice secondary effects not included in the model such as
saturation or clipping will provide an effective limit.

Finally, note that the cost of providing authentication is readily
apparent since the inner bound from \eqref{eq:d2:lb} represents the
distortions achievable when the reconstruction need not be authentic.
Since for a fixed SNR, the bounds converge asymptotically for large
$D_e$, and for a fixed $D_e \geq \sigma_N^2$ the bounds converge
asymptotically for large SNR, we conclude that the price of
authentication is negligible in these regimes.  However, for low
$\Diste$ regimes of operation, requiring authenticity strictly reduces
the achievable distortion region.  This behavior is analogous to that
observed in the binary-Hamming case.

\section{Comparing Authentication Architectures}
\label{sec:discussion}

The most commonly studied architectures for authentication are robust
watermarking (i.e., self-embedding) and fragile watermarking.  In the
sequel we compare these architectures to that developed in this paper.

\subsection{Authentication Systems Based on Robust Watermarking}
\label{sec:robust}

The robust watermarking approach to encoding for authentication (see,
e.g., \cite{schneider_1996, queluz, bat_kut, rey_2000, Lin_2001})
takes the form of a quantize-and-embed strategy.  The basic steps of
the encoding are as follows.  First, the source $\ful{S}$ is quantized
to a representation in terms of bits using a source coding
(compression) algorithm.  Second the bits are protected using a
cryptographic technique such as a digital signature or hash function.
Finally, the protected bits are embedded into the original source
using an information embedding (digital watermarking) algorithm.  At
the decoder, the embedded bits are extracted.  If their authenticity
is verified via the appropriate cryptographic technique, a
reconstruction of the source is produced from the bits.  Otherwise,
the decoder declares that an authentic reconstruction is not possible.

It is straightforward to develop the information-theoretic limits of
such approaches, and to compare the results to the optimum systems
developed in the preceding sections.  In particular, if we use optimum
source coding and information embedding in the quantize-and-embed
approach, it follows that, in contrast to Theorem~\ref{th:main}, the
distortion pair $(\Diste,\Distr)$ lies in the achievable distortion
region for a quantize-and-embed structured solution to the problem
\eqref{eq:authprob} if and only if there exists distributions
$p(\nsrch|\nsrc)$ and $p(\naux|\nsrc)$, and a function
$\senc{\cdot,\cdot}$, such that
\begin{subequations}
\label{eq:qe}
\begin{align}
I(\nAux;\nChOut) - I(\nSrc;\nAux)  &\geq  I(\nSrc;\nSrch) \label{eq:qe:a} \\
E[\diste(\nSrc,\senc{\nAux,\nSrc})]  &\leq  \Diste \label{eq:qe:b} \\
E[\distr(\nSrc,\nSrch)]  &\leq  \Distr. \label{eq:qe:c}
\end{align}
These results follow from the characterization of the rate-distortion
function of a source \cite{cover} and the capacity of information
embedding systems with distortion constraints as developed in
\cite{rjb_bc_gw_preprint} as an extension of \cite{gelfand_1980}.
\end{subequations}

Comparing \eqref{eq:qe} to \eqref{eq:thm} with $\nSrch=g(\nAux)$ we
see that quantize-and-embed systems are unnecessarily constrained,
which translates to a loss of efficiency relative to the optimum joint
source--channel--authentication coding system constructions of
\secref{sec:proofs}.  This performance penalty can be quite severe in
the typical regimes of interest, as we now illustrate.  In particular,
we quantify this behavior in the two example scenarios considered
earlier: the binary-Hamming and Gaussian-quadratic cases.

\subsubsection{Example: Binary-Hamming Case}

\begin{figure*}[tbp]
\centering
\psfrag{D1}{\Large$\Diste$}
\psfrag{D2}{\Large$\Distr$}
\psfrag{TT1}{$\crossProb=0.05$}
\psfrag{TT2}{$\crossProb=0.10$}
\psfrag{TT3}{$\crossProb=0.15$}
\psfrag{TT4}{$\crossProb=0.20$}
\includegraphics[angle=0]{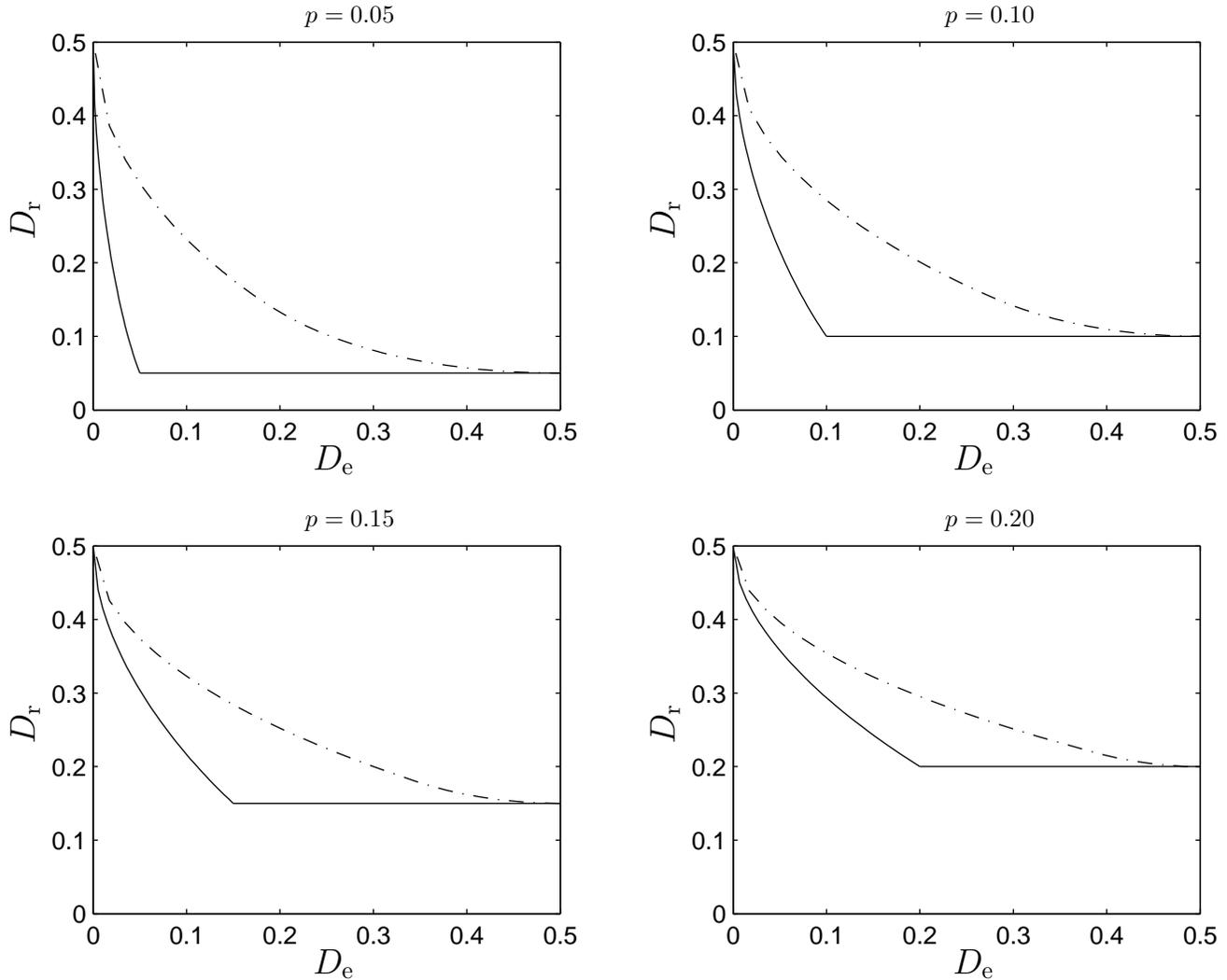}
\caption{Performance loss of quantize-and-embed systems for the
Binary-Hamming scenario with various reference channel crossover
probabilities $p$.  The solid curve depicts the boundary of the
achievable regions for the optimum system; the dashed curve depicts
that of the best quantize-and-embed system. \label{fig:h_comp}}
\end{figure*}

In this scenario, the rate-distortion function is \cite{cover} 
\begin{equation} 
R(\Distr) = 1 - h(\Distr),
\label{eq:rd-bh}
\end{equation}
while the information embedding capacity is (see
\cite{rjb_bc_gw_preprint}) the upper concave envelope of the function
\begin{equation} 
g_p(\Diste) =
  \begin{cases}
    0, & \text{if $0\leq d<p$,} \\
    h(\Diste)-h(p), & \text{if $p\leq \Diste\leq1/2$,}
  \end{cases}
\label{eq:g-ie}
\end{equation}
i.e., 
\begin{equation} 
C(\Diste) =
  \begin{cases}
   \displaystyle 
   \frac{g_p(\Dist_p)}{\Dist_p}\Diste, & \text{if $0\leq \Diste\leq \Dist_p$,} \\
	g_p(\Diste), & \text{if $\Dist_p< \Diste \leq 1/2$,}
  \end{cases} 
\label{eq:ie-bh}
\end{equation}
where $\Dist_p=1-2^{-h(p)}$.  Equating $R$ in \eqref{eq:rd-bh} to $C$
in \eqref{eq:ie-bh}, we obtain a relation between $\Distr$ and
$\Diste$.  This curve is depicted in Fig.~\ref{fig:h_comp} for
different reference channel parameters.  As this figure reflects, the
optimum quantize-and-embed system performance lies strictly inside the
achievable region for the binary-Hamming scenario developed in
\secref{sec:binary_hamming}, with the performance gap largest for the
cleanest reference channels.  Moreover, since as we saw in
Section~\ref{sec:exdr} clean reference channels correspond to ensuring
small encoding and reconstruction distortions, this means that
quantize-and-embed systems suffer the largest losses precisely in the
regime one would typically want to operate in.

\subsubsection{Example: Gaussian-Quadratic Case}

\begin{figure*}[hbt]
\centering
\psfrag{-10DBSNR}{\Large\hspace{.5in} -10 dB SNR}
\psfrag{0DBSNR}{\Large\hspace{.5in} 0 dB SNR}
\psfrag{10DBSNR}{\Large\hspace{.5in} 10 dB SNR}
\psfrag{30DBSNR}{\Large\hspace{.5in} 30 dB SNR}
\psfrag{DAX1}{\Large\hspace{-30pt}\raisebox{-.05in}{$\Diste/\sigma_N^2$ (in dB)}}
\psfrag{DAX2}{\Large\hspace{-30pt}\raisebox{.1in}{$\Distr/\sigma_N^2$ (in dB)}}
\includegraphics[angle=0]{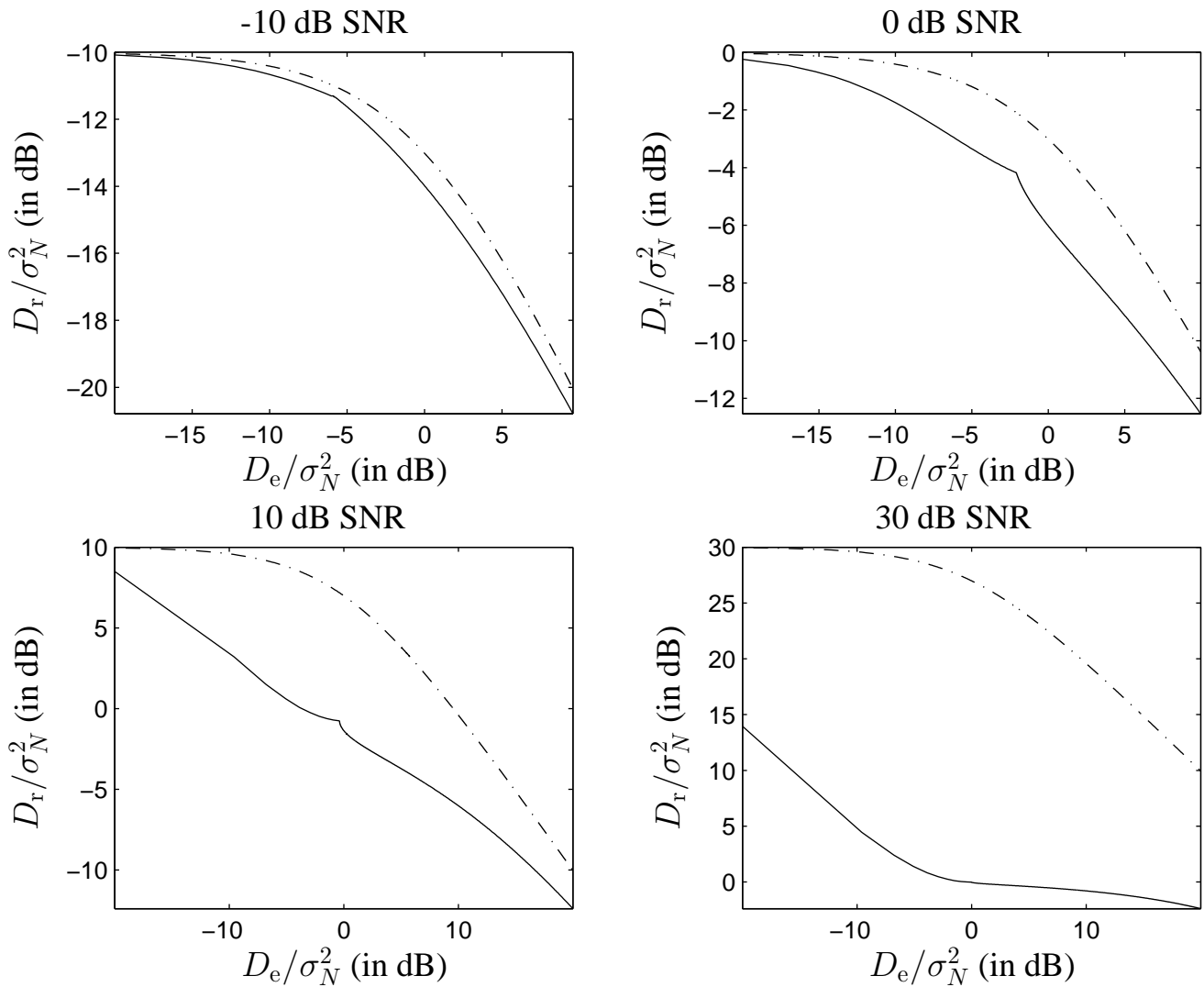}
\caption{Performance loss of quantize-and-embed systems for the
Gaussian-quadratic scenario at various reference channel SNRs.  The
solid curve depicts the asymptotic outer bound of the
achievable regions for the optimum system; the dashed curve depicts
that of the best quantize-and-embed system. \label{fig:g_comp}}
\end{figure*}

In this scenario, the rate-distortion function is \cite{cover}
\begin{equation} 
R(\Distr) = 
\begin{cases} \frac{1}{2} \log\frac{\sigma_{\nSrc}^2}{\Distr}, 
                & 0 \le \Distr \le \sigma_{\nSrc}^2 \\
              0, & \Distr > \sigma_{\nSrc}^2,
\end{cases}
\label{eq:rd-gq}
\end{equation}
while the information embedding capacity is \cite{costa_83}
\begin{equation} 
C(\Diste) = \frac{1}{2}\log\left(1 + \frac{\Diste}{\sigma_N^2}\right).
\label{eq:ie-gq}
\end{equation}
Again, equating $R$ in \eqref{eq:rd-gq} to $C$ in \eqref{eq:ie-gq}, we
obtain the following relation between $\Distr$ and $\Diste$ for all
$\Diste > 0$:
\begin{equation}
\Distr =\frac{\sigma_{\nSrc}^2}{(1 + \Diste/\sigma_N^2)}.
\label{eq:qe-gq}
\end{equation}
This curve is depicted in Fig.~\ref{fig:g_comp} for different
reference channel SNRs.  This figure reflects that the optimum
quantize-and-embed system performance lies strictly inside the
achievable region for the Gaussian-quadratic scenario developed in
\secref{sec:gaussian}.  Likewise, the performance gap is largest for
the highest SNR reference channels.  Indeed, comparing the inner bound
\eqref{eq:d2:lb} on the performance of the optimum system with that of
quantize-and-embed, i.e., \eqref{eq:qe-gq}, we see that while
quantize-and-embed incurs no loss at low SNR:
\begin{equation} 
\frac{\Distr^{\mathrm{qe}}}{\Distr} \rightarrow 1 \quad\text{as}\quad
\frac{\sigma_{\nSrc}^2}{\sigma_N^2} \rightarrow 0,
\end{equation}
at high SNR the loss is as much as $\mathrm{SNR}/2$ for
$\Diste\ge\sigma_N^2$:
\begin{equation} 
\frac{\sigma_N^2}{\sigma_{\nSrc}^2}
\frac{\Distr^{\mathrm{qe}}}{\Distr} \rightarrow
\frac{1}{1+\Diste/\sigma_N^2} \le \frac{1}{2} \quad\text{as}\quad
\frac{\sigma_{\nSrc}^2}{\sigma_N^2} \rightarrow \infty,
\end{equation}
where we have used $\Distr^{\mathrm{qe}}$ to denote the
quantize-and-embed reconstruction distortion \eqref{eq:qe-gq}.

Hence, as in the binary-Hamming case, we see again that
quantize-and-embed systems suffer the largest losses in the regime
where one is most interested in operating --- that where the editor is
allowed to make only perturbations small enough that the corresponding
encoding and reconstruction distortions are small.\footnote{It should
be emphasized that while one could argue that the quadratic distortion
measure is a poor measure of semantic proximity in many applications,
such reasoning confuses two separate issues.  We show here that
quantize-and-embed systems are quite poor when the quadratic measure
corresponds \emph{exactly} to the semantics of interest.  For problems
where it is a poor match, one can expect systems based on more
accurate measures to exhibit the same qualitative behavior --- that
quantize-and-embed systems will be least attractive in regimes where
the source encodings and reconstructions are constrained to be
semantically close to the original source.}

\subsection{Authentication Systems Based on Fragile Watermarking}
\label{sec:fragile}

A fundamentally different approach to the authentication problems of
this paper is based on constraining the semantic severity of the
modifications the editor is allowed to make.  In particular, given a
distortion measure that captures the semantic impact of edits to the
content, the decoder will declare the edited content authentic if and
only if the distortion is below some predetermined threshold.  We
refer to these as authentication systems based on semantic
thresholding.

It is important to appreciate that the manner in which the editor is
constrained in systems based on semantic thresholding is qualitatively
quite different from the way the editor is constrained in the systems
developed in this paper.  In particular, in our formulation, the
editor is contrained according to a reference channel model that can
be freely chosen --- independently of any semantic model.

While in this section we are primarily interested in discussing the
properties of such systems, we first briefly describe how such
systems can be designed.  We begin by noting that role of the encoder
in such systems is to mark the original content so as to enable the
eventual decoder to estimate the distortion between the edited content
and that original content, despite not having direct access to the
latter.

One approach to such a problem would be to use the self-embedding idea
discussed in \secref{sec:robust}.  In particular, a compressed version
of the original content would be embedded into that content so that it
could be reliably extracted from the edited content by the decoder and
used in the distortion calculation.  In practice, such self-embedding
can be somewhat resource inefficient, much as it was in the context of
\secref{sec:robust}.  Instead, an approach based on so-called fragile
watermarking is more typically proposed, which allows the decoder to
measure the distortion without explicitly being given an estimate of
the original content.  With this approach, distortion in the known
watermark that results from editing the content are used to infer the
severity of distortion in the content itself.

Typical implementations of the fragile watermarking approach to
encoding for authentication (see, e.g., \cite{kundur, yeung_1997,
wolfgang_1996, eggers_2001}) take the following form.  A watermark
message $M$ known only to the encoder and decoder (and kept secret
from the editor) is embedded into the source signal by the encoder.
The editor's processing of the encoded content indirectly perturbs the
watermark.  A decoder extracts this perturbed watermark $\hat{M}$,
measures the size of the perturbation (e.g., by computing the
distortion between $\hat{M}$ and $M$ with respect to some suitable
measure), then uses the result to assess the (semantic) severity of
the editing the content has undergone.  If the severity is below some
predetermined threshold, the decoder declares the signal to be
authentic.

A detailed information-theoretic characterization of authentication
systems based on semantic thresholding is beyond the scope of this
paper.  However, in the sequel we emphasize some important qualitative
differences in the security characteristics between such schemes and
those developed in this paper.  In particular, as we now develop,
there is a fundamental vulnerability in semantic thresholding schemes
that results from their inherent sensitivity to mismatch in the chosen
semantic model.

To see this, consider a mismatch scenario in which the authentication
system is designed with an incorrect semantic model (distortion
measure).  If the system is based on semantic thresholding, then an
attacker who recognizes the mismatch can exploit this knowledge to
make an edit that is semantically significant, but which the system
will deem as semantically insignificant due to the model error, and
thus accept as authentic.  Thus, for such systems, a mismatch can lead
to a security failure.

By contrast, for the authentication systems developed in this paper,
designing the system based on the incorrect semantic model reduces the
efficiency of the system, but does not impact its security.  In
particular, use of the incorrect semantic model leads to encodings
and/or authentic reconstructions with unnecessarily high distortions
(with respect to the correct model).  However, attackers cannot
exploit this to circumvent the security mechanism, since they are
constrained by the reference channel, which is independent of the
semantic model.

From such arguments, one might conclude that systems based on semantic
thresholding might be preferable so long as care is taken to develop
accurate semantic models.  However, such a viewpoint fails to
recognize that in practice some degree of mismatch is inevitable ---
the high complexity of accurate semantic models makes them inherently
difficult to learn.  Thus, in a practical sense, authentication
systems based on semantic thresholding are intrinsically less secure
than those developed in this paper.

\section{Layered Authentication: Broadcast Reference Channels}
\label{sec:layered}

For many applications, one might be interested in an authentication
system with the property that an authentic reconstruction is always
produced, but that its quality degrades gracefully with the
extensiveness of the editing the content has undergone.  In this
section we show that discretized versions of such behavior are
possible, and can be built as a natural extension of the formulation
of this paper.

To develop this idea, we begin by observing that the systems developed
thus far in the paper represent a first-order approximation to such
behavior.  In particular, for edits consistent with the reference
channel model, an authentic reconstruction of fixed quality is
produced.  When the editing is not consistent with the reference
channel, the only possible authentic reconstruction is the minimal
quality one one obtained from the \emph{a priori} distribution for the
content, since the edited version must be ignored altogether.  In this
section, we show that by creating a hierarchy of reference channels
corresponding to increasing amounts of editing, one can create
multiple authentication reconstructions.  In this way, a graceful
degradation characteristic can be obtained to any desired granularity.

Such systems can be viewed as layered authentication systems, and
arise naturally out of the use of broadcast reference channel models.
With such systems there is a fixed encoding of the source that incurs
some distortion.  Then, from edited content that is consistent with
any of the constituent reference channels in the broadcast model, the
decoder produces an authentic reconstruction of some corresponding
fidelity.  Otherwise, the decoder declares that an authentic
reconstruction is not possible.

For the purpose of illustration, we focus on the two-user memoryless
degraded broadcast channel \cite{cover} as our reference channel.
This corresponds to a two-layer authentication system.  For
convenience, we refer to the strong channel as the ``mild-edit'' one,
and the weak channel, which is a degraded version of the strong one,
as the ``harsh-edit'' one.  Edits consistent with the mild-edit branch
of the reference channel will allow higher quality authentic
reconstructions, which we will call ``fine,'' while edits consistent
with the harsh-edit branch will allow lower quality authentic
reconstructions, which we will call ``coarse''.  For edits
inconsistent with either branch, the only authentic reconstruction
will be one that ignores the edited data, which will be of lowest
quality.

In this scenario, for any prescribed level of encoding distortion
$\Diste$, there is a fundamental trade-off between the achievable
distortions $\Distrf$ and $\Distrc$ of the corresponding fine and
coarse authentic reconstructions, respectively.  Of course
$\Distrc\ge\Distrf$ will always be satisfied.  However, as we will see,
achieving smaller values of $\Distrc$ in general requires accepting
larger values of $\Distrf$ and vice-versa.  Using the ideas of this
paper, one can explore the fundamental nature of such trade-offs.

\subsection{Achievable Distortion Regions}

\begin{figure*}[tbp]
\centering
\psfrag{X}{$\ful{\nSrc}$}
\psfrag{Y}{$\ful{\nChIn}$}
\psfrag{A}{$\ful{\nChOutDeg}$}
\psfrag{B}{$\ful{\nChOutRef}$}
\psfrag{C}{$\ful{\nSrch_{\mathrm{c}}}$}
\psfrag{D}{$\ful{\nSrch_{\mathrm{f}}}$}
\includegraphics[angle=0,width=6in]{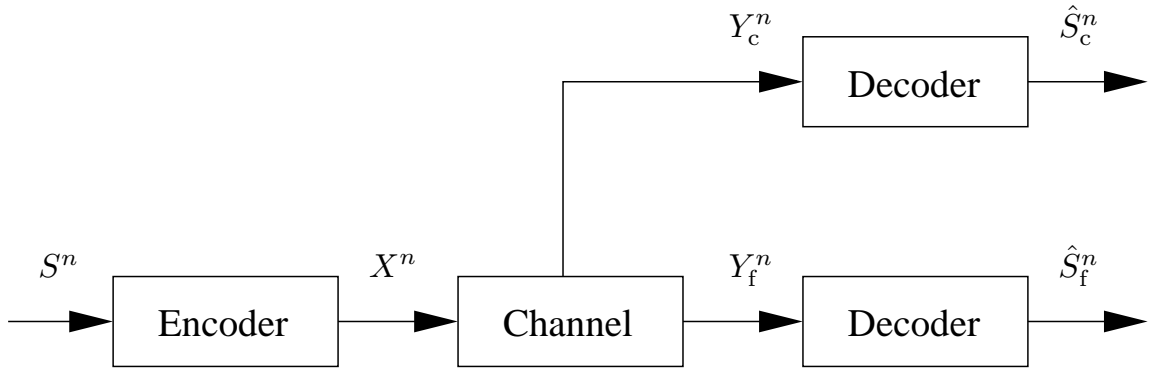}
\caption{Two-layer authentication system operation when the reference
channel is in effect.  From the outputs $\nChOutRef$ and $\nChOutDeg$
of the degraded broadcast reference channel, corresponding to mild and
harsh editing, the respective fine and coarse authentic
reconstructions $\ful{\nSrch_{\mathrm{f}}}$ and
$\ful{\nSrch_{\mathrm{c}}}$ are produced.  The common encoding
obtained from the source $\ful{\nSrc}$ is $\ful{\nChIn}$.
\label{fig:broadcast_prob_mod}}
\end{figure*}

The scenario of interest is depicted in
Fig.~\ref{fig:broadcast_prob_mod}.  As a natural generalization of its
definition in the single-layer context \eqref{eq:authprob}, an
instance of the layered authentication problem consists of the eight-tuple
\begin{equation} 
\left\{ \srcAlph, p(\nsrc), \chinAlph, \choutAlph,
        p(\nchoutdeg|\nchoutref), p(\nchoutref|\nchin), 
        \diste(\cdot,\cdot), \distr(\cdot,\cdot) \right\},
\label{eq:layauthprob}
\end{equation}
where, since our reference channel is a degraded broadcast channel,
the reference channel law takes the form
\begin{equation}
\label{eq:deg_cond}
p(\ful{\nchoutdeg},\ful{\nchoutref}|\ful{\nchin}) =
p(\ful{\nchoutdeg}|\ful{\nchoutref})\,p(\ful{\nchoutref}|\ful{\nchin}).
\end{equation}

Let $\ful{\nSrch_{\mathrm{c}}}$ denote the (coarse) authentic
reconstruction obtained when decoder input is consistent with the
harsh-edit output of the reference channel, and let
$\ful{\nSrch_{\mathrm{f}}}$ denote the (fine) authentic reconstruction
obtained when decoder input is consistent with the mild-edit output of
the reference channel.  In turn, the corresponding two reconstruction
distortions are defined according to
\begin{subequations}
\begin{align} 
\Distrc &= \frac{1}{n} \sum_{i=1}^n
\distrc(\ful{\nSrc},\ful{\nSrch_{\mathrm{c}}})\\
\Distrf &= \frac{1}{n} \sum_{i=1}^n
\distrf(\ful{\nSrc},\ful{\nSrch_{\mathrm{f}}}).
\end{align}
\label{eq:distr-lay}
\end{subequations}

The following theorem develops trade-offs between the encoding
distortion $\Diste$, and the two reconstruction distortions
\eqref{eq:distr-lay} that are achievable.  
\begin{theorem}
\label{thm:layered}
The distortion triple $(\Diste, \Distrc, \Distrf)$ lies in the
achievable distortion region for the layered authentication problem
\eqref{eq:layauthprob} if there exist distributions
$p(\nauxdeg,\nauxref|\nsrc)$ and $p(\nchin|\nauxdeg,\nauxref,\nsrc)$,
and functions $\degsedec{\cdot}$ and $\refsedec{\cdot}{\cdot}$ such
that
\begin{subequations}
\label{eq:layered:thm}
\begin{align}
I(\nAuxDeg;\nChOutDeg) - I(\nSrc;\nAuxDeg) &\geq 0 \label{eq:layered:a}\\
I(\nAuxRef;\nChOutRef|\nAuxDeg) - I(\nSrc;\nAuxRef|\nAuxDeg)  &\geq  0
\label{eq:layered:b} \\
E[\diste(\nSrc,\nChIn)]  &\leq  \Diste  \label{eq:layered:c}\\
E[\distrc(\nSrc,\degsedec{\nAuxDeg})] &\leq \Distrc. \label{eq:layered:d}\\
E[\distrf(\nSrc,\refsedec{\nAuxDeg}{\nAuxRef})]  &\leq
\Distrf. \label{eq:layered:e} 
\end{align}
\end{subequations}
\end{theorem}
In this theorem, the achievable distortion region is defined in a
manner that is the natural generalization of that for single-layer
systems as given in Definition~\ref{def:adr}.

In the interests of brevity and since it closely parallels that for
the single-layer case, we avoid a formal derivation of this result.
Instead, we sketch the key ideas of the construction.  We also leave
determining the degree to which the distortion region can be further
extended via more elaborate coding for future work.

\begin{proof}[Sketch of Proof:]

First a codebook $\cbookdeg$ is created for the harsh-edit layer at
rate $\cbkRdeg = I(\nAuxDeg;\nSrc) + 2\gamma$ where only
$2^{n(\cbkRdeg+\gamma)}$ codewords are marked as admissible as in
\thrmref{th:main}.  Then for each codeword $\cdeg \in \cbookdeg$ an
additional random codebook $\cbookref(\cdeg)$ of rate $\cbkRref =
I(\nAuxRef;\nSrc|\nAuxDeg) + 2\gamma$ is created according to the
marginal distribution $p(\nauxref|\nauxdeg)$ where only $2^{n(\cbkRref
+ \gamma)}$ codewords are marked as admissible.

The encoder first searches $\cbookdeg$ for an admissible codeword $\cdeg$
jointly typical with the source and then searches $\cbookref(\cdeg)$ for a
refinement $\cref$ that is jointly typical with the source.  The pair
$(\cdeg,\cref)$ is then mapped into the channel according to
$p(\nchin|\nauxdeg,\nauxref,\nsrc)$.  By standard arguments the
encoding will succeed with high probability provided that $\cbkRdeg >
I(\nAuxDeg;\nSrc)$ and $\cbkRref > I(\nAuxRef;\nSrc|\nAuxDeg)$.  

When the channel output is consistent with either output of the
reference channel, the decoder locates an admissible codeword $\cdegh
\in \cbookdeg$ jointly typical with the signal.  If the
signal is consistent with the harsh-edit output of the reference
channel, in particular, the decoder then produces the coarse authentic
reconstruction $\ful{\nSrch}_{\mathrm{c}} = \degsedec{\cdegh}$.
However, if the signal is consistent with the mild-edit output of
the reference channel, the decoder then proceeds to locate an
admissible $\crefh \in \cbookref(\cdegh)$ and produces the fine
authentic reconstruction $\ful{\nSrch}_{\mathrm{f}} =
\refsedec{\cdegh}{\crefh}$.

By arguments similar to those used in the single-layer case (i.e.,
proof of \thrmref{th:main}), this strategy achieves vanishingly small
probabilities of successful attack, and when the reference channel is
in effect meets the distortion targets provided that $\cbkRdeg <
I(\nAuxDeg;\nChOutDeg)$ and $\cbkRref <
I(\nAuxRef;\nChOutRef|\nAuxDeg)$.

\end{proof}

\subsection{Example: Gaussian-Quadratic Case}

The Gaussian-quadratic case corresponds to the mild- and harsh-edit
outputs of the reference channel taking the forms $\nChOutRef = \nChIn
+ N$ and $\nChOutDeg = \nChOutRef + V$, respectively, where $N$ and
$V$ are Gaussian random variables independent of each other, as well
as $\nSrc$ and $\nChIn$.

\begin{figure*}[tbp]
\centering
\psfrag{Sn}{\large$\ful{\nSrc}$}
\psfrag{Xn}{\large$\ful{\nChIn}$}
\psfrag{Un}{\large$\ful{\nAuxDeg}$}
\psfrag{Tn}{\large$\ful{\nAuxRef}$}
\psfrag{Wna}{\large$\ful{\nChOutRef}$}
\psfrag{Wnb}{\large$\ful{\nChOutDeg}$}
\psfrag{Wnc}{\large$\ful{Z}$}
\includegraphics[angle=0,width=5in]{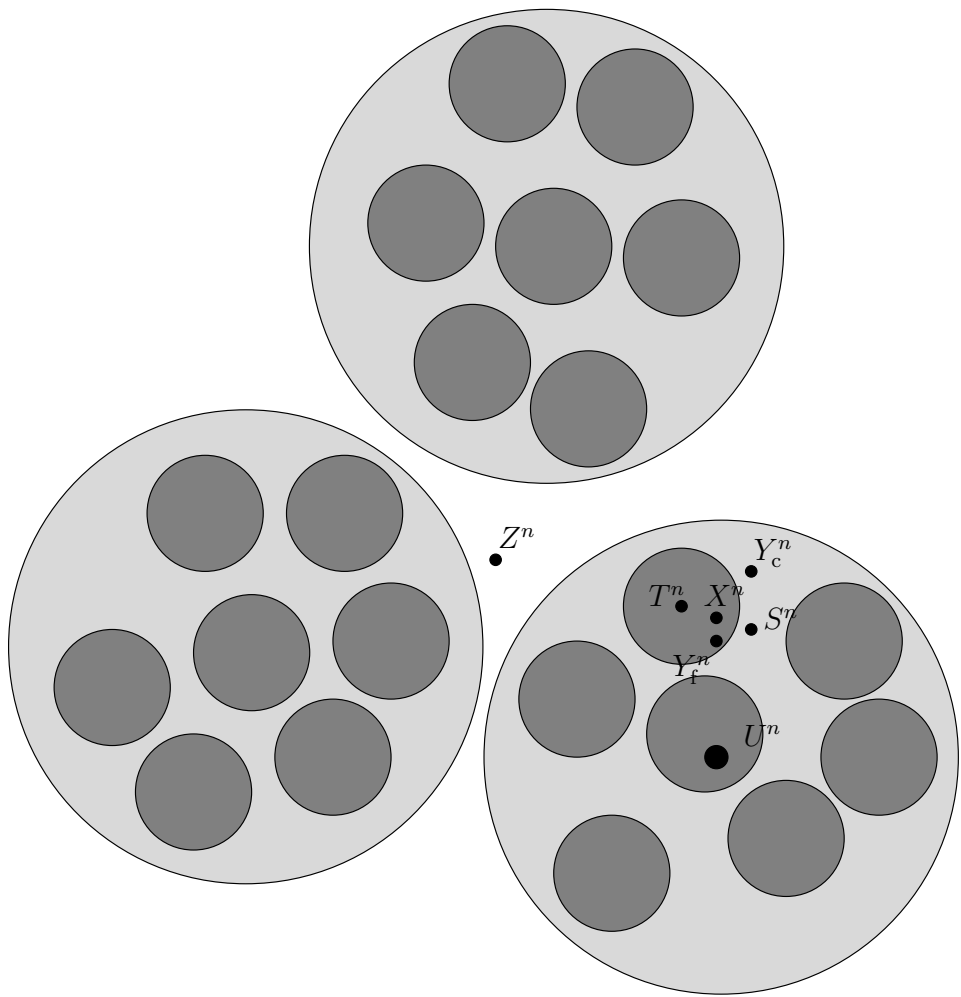}
\caption{Illustration of the nested codebook geometry associated with
a two-layer authentication system for the Gaussian-quadratic scenario.
The centers of large and small shaded spheres correspond to admissible
coarse and fine authentic reconstructions, respectively.
\label{fig:sigspacelay}}
\end{figure*}

For this case, a natural approach to the layered authentication system
design has the structure depicted in Fig.~\ref{fig:sigspacelay}, which
generalizes that of the single-layer systems developed in
Section~\ref{sec:gaussian}.  The encoder determines the codeword
$\ful{\nAuxRef}$ nearest the source $\ful{\nSrc}$, then perturbs
$\ful{\nAuxRef}$ so as to reduce the encoding distortion, producing
the encoding $\ful{\nChIn}$.  If the channel output stays within the
darkly shaded sphere centered about $\ful{\nAuxRef}$, e.g., producing
$\ful{\nChOutRef}$ as shown, the decoder produces a fine-grain
authentic reconstruction from $\ful{\nAuxRef}$.  If the channel output
is outside the darkly shaded sphere, but inside the encompassing
lightly shaded sphere centered about $\ful{\nAuxDeg}$, e.g., producing
$\ful{\nChOutDeg}$ as shown, the decoder produces a coarse-grain
authentic reconstruction from $\ful{\nAuxDeg}$.  If the channel output
is outside any shaded region, e.g., producing $\ful{Z}$, the decoder
indicates that an authentic reconstruction is not possible.

An achievable distortion region for this layered authentication
scenario is obtained from Theorem~\ref{thm:layered} with the
auxiliary random variables chosen according to
\begin{align}
\nAuxDeg &= \nSrc + A/\alpha\\
\nAuxRef &= \nSrc + B/\beta\\
\nChIn   &= \nSrc + A + B.
\end{align}
where $A$ and $B$ are Gaussian random variables independent of
$\nSrc$.  Choosing $\degsedec{\cdot}$ and $\refsedec{\cdot}{\cdot}$ to
be the minimum mean-square error estimates of $\nSrc$ from $\nAuxDeg$
and $(\nAuxDeg,\nAuxRef)$, respectively, yields
\begin{align}
\Diste &= \sigma_A^2 + \sigma_B^2\\
\Distrc &= \sigma_{\nSrc}^2\left(1 -
\frac{E[\nSrc\nAuxDeg]^2}{E[\nSrc^2]E[\nAuxDeg^2]}\right) =
\frac{\sigma_{\nSrc}^2\sigma_A^2}{\sigma_A^2 + \alpha^2
\sigma_{\nSrc}^2}\\
\Distrf &= \sigma_{\nSrc}^2 - \Lambda_{\nSrc,[\nAuxDeg \nAuxRef]}
\Lambda_{[\nAuxDeg \nAuxRef]}^{-1} \Lambda_{[\nAuxDeg \nAuxRef],
  \nSrc} \notag\\
& =
\frac{\sigma_{\nSrc}^2\sigma_A^2\sigma_B^2}{\beta^2 \sigma_{\nSrc}^2
\sigma_A^2 + \sigma_A^2\sigma_B^2 + \alpha^2\sigma_{\nSrc}^2\sigma_B^2},
\end{align}
where $\Lambda$ with a single subscript denotes the covariance of its
argument, and $\Lambda$ with a subscript pair denotes the
cross-covariance between its arguments.

To produce $\ful{\nSrch_{\mathrm{c}}}$, a decoder essentially views
$B$ as additive channel noise.  Therefore, we can immediately apply
the arguments from \secref{sec:gaussian:ach_dist} to obtain
\begin{multline}
I(\nAux;\nChOutDeg)-I(\nSrc;\nAux) =\\
\frac{1}{2}\log\frac{\sigma_{A}^2(\sigma_{A}^2 +
\sigma_{\nSrc}^2 +
\sigma_N^2+\sigma_V^2+\sigma_B^2)}{\sigma_{A}^2\sigma_{\nSrc}^2(1-\alpha)^2 +
(\sigma_N^2+\sigma_V^2+\sigma_B^2)(\sigma_{A}^2 +
\alpha^2\sigma_{\nSrc}^2)}.
\end{multline}
From this we can solve for $\alpha$ as in the single-layer case of
\secref{sec:gaussian:ach_dist} by simply replacing $\sigma_{\genrv}^2$
and $\sigma_{N}^2$ with $\sigma_{A}^2$ and $\sigma_N^2 + \sigma_V^2 +
\sigma_B^2$, respectively, in \eqref{eq:rp-def}.

\begin{figure*}[tbp]
\centering
\psfrag{D2AXLABEL}{\raisebox{-.05in}{\Large$\Distrc/\sigma_N^2$ (in dB)}}
\psfrag{D3AXLABEL}{\raisebox{-.025in}{\Large$\Distrf/\sigma_N^2$ (in dB)}}
\includegraphics[angle=0,width=6in]{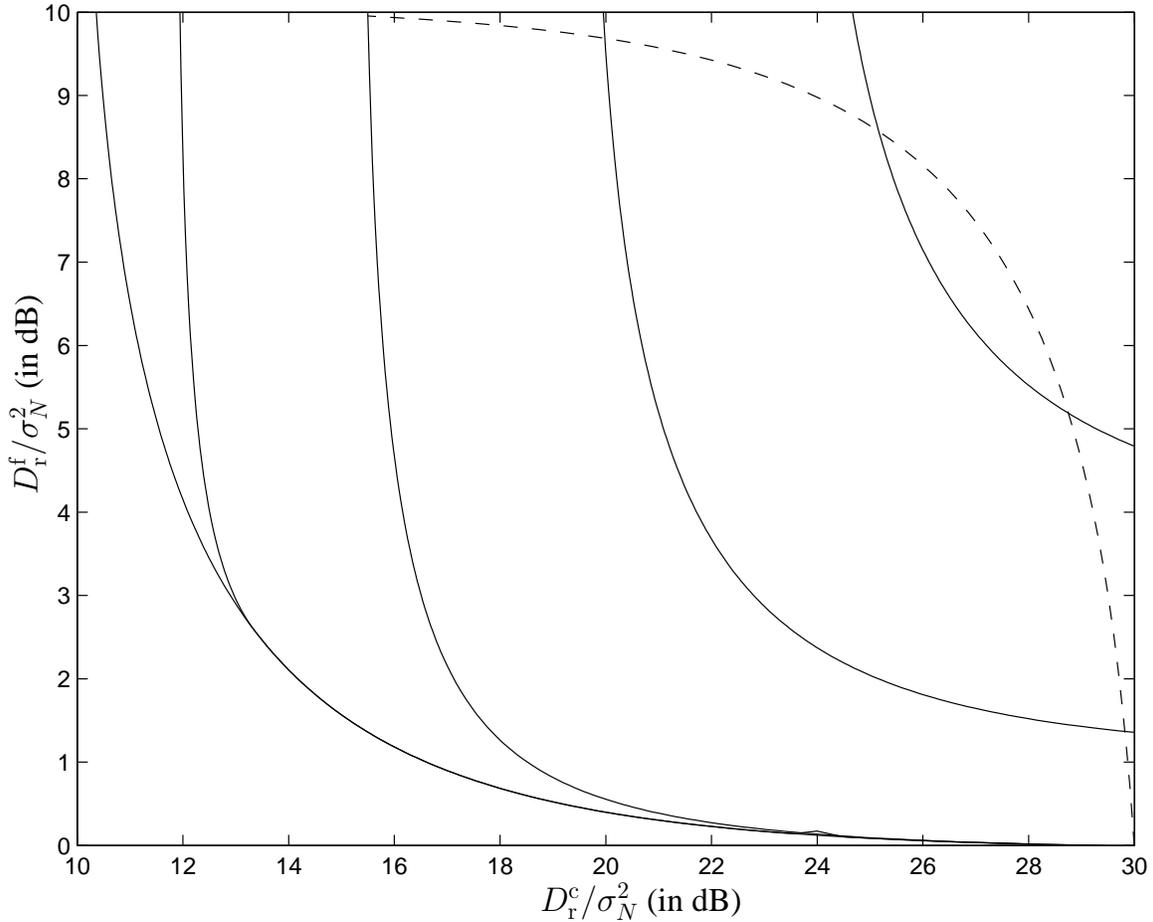}
\caption{Achievable fine and coarse quality reconstruction distortion
pairs $(\Distrf,\Distrc)$ in a layered authentication system for the
Gaussian-quadratic case with $\sigma_{\nSrc}^2/\sigma_N^2 = 30$ dB,
$\sigma_V^2/\sigma_N^2 = 10$ dB, and $\sigma_N^2 = 1$.  From left to
right, the curves are the boundaries of achievable distortion regions
corresponding to encoding distortions of $\Diste/\sigma_N^2 = 10, 5, 0,
-5, -10$ dB.  The dashed curve corresponds to time-sharing between two
operating points on the $\Diste/\sigma_N^2=0$ dB
curve. \label{fig:layered_g_plots}}
\end{figure*}

Finally, since
\begin{multline}
I(\nSrc;\nAuxRef|\nAuxDeg) - I(\nAuxRef;\nChOutRef|\nAuxDeg)
=
H(\nAuxRef|\nAuxDeg,\nChOutRef) - H(\nAuxRef|\nAuxDeg,\nSrc) \\
=H(\nAuxRef,\nAuxDeg,\nChOutRef)-H(\nAuxDeg,\nChOutRef)\\ -
H(\nAuxRef,\nAuxDeg,\nSrc)+ H(\nAuxDeg,\nSrc).
\end{multline}
we see that \eqref{eq:layered:b} implies
\begin{equation} 
\frac{\det(\Lambda_{[\nAuxRef \nAuxDeg
\nChOutRef]})}{\det(\Lambda_{[\nAuxDeg \nChOutRef]})} \leq 
\frac{\det(\Lambda_{[\nAuxRef \nAuxDeg \nSrc]})}{\det(\Lambda_{[\nAuxDeg \nSrc]})}.
\label{eq:det_cond}
\end{equation}
By varying $\sigma_A^2$, $\sigma_B^2$, and $\beta$ such that
\eqref{eq:det_cond} is satisfied we can trace out the volume of an
achievable distortion region.  Fig.~\ref{fig:layered_g_plots} shows
slices of this three dimensional region by plotting the fine and
coarse reconstruction distortions $\Distrf$ and $\Distrc$ for various
values of the encoding distortion $\Diste$.  Note that it follows from
our single-layer inner bounds that for a particular choice of encoding
distortion $\Diste$, the achievable trade-offs between $\Distrc$ and
$\Distrf$ are contained within the region
\begin{align}
\Distrc &\ge \frac{\sigma_\nSrc^2 (\sigma_N^2+\sigma_V^2)}{\sigma_N^2
+ \sigma_V^2 + 
\left(\sqrt{\Diste} + \sigma_\nSrc\right)^2} \label{eq:Distrc-bd}\\ 
\Distrf &\ge \frac{\sigma_\nSrc^2\sigma_N^2}{\sigma_N^2 + \left(\sqrt{\Diste} + \sigma_\nSrc\right)^2},\label{eq:Distrf-bd}
\end{align}
where obviously the lower bound of \eqref{eq:Distrf-bd} is smaller than
that of \eqref{eq:Distrc-bd}.  

A simple alternative to the layering system for such authentication
problems is time-sharing, whereby some fraction of time the encoder
uses a codebook appropriate for the harsh-edit reference channel, and for
the remaining time uses a codebook appropriate for the mild-edit reference
channel.  When the harsh-edit reference channel is in effect, the decoder
produces the coarse authentic reconstruction for the fraction of time
the corresponding codebook is in effect and produces zero the rest of
the time.  When the mild-edit reference channel is in effect, the decoder
produces the fine authentic reconstruction during the fraction of time
the corresponding codebook is in effect, and produces the coarse
reconstruction for the remaining time (since the broadcast channel is
a degraded one).  However, as Fig.~\ref{fig:layered_g_plots}
also illustrates, this approach is in general quite
inefficient: the use of such time-sharing results in a substantially
smaller achievable region.

\section{Concluding Remarks}
\label{sec:conc}

This paper develops one meaningful formulation for authentication
problems in which the content may undergo a variety of types of
legitimate editing prior to authentication.  As part of this
formulation, we adopt a particular formal notion of security in such
settings.  For such a formulation, and with the simplest classes of
models, we establish that secure authentication systems can be
constructed, and subsquently analyze their fundamental performance
limits.  From these models, we further develop how such systems offer
significant advantages over other proposed solutions.

Many opportunities for further research remain.  For example,
extensions of the main results to richer content, semantic, and edit
models may provide additional insights into the behavior of such
sysems.  It would also be useful to understand the degree to which
robust and/or universal solutions exist for the problem; such
approaches seek to avoid requiring accurate prior model knowledge
during system design.

There are additional opportunities to further refine the analysis even
for the existing models.  For example, characterizing the manner in
which asymptotic limits are approached --- for example via error
exponents --- would provide useful engineering insights.  Likewise,
further analyzing public-key formulations, in which edits are more
generally subject to computational constraints, could also be
revealing.  From this persective, the Appendix represents but a
starting point.

More generally, identifying and relating other meaningful notions of
security for such problems would be particularly useful in putting the
results of this paper in perspective.  For example, a broader unifying
framework for characterizing and comparing different notions of
security could provide a mechanism for selecting a formulation best
matched to the social needs and/or engineering constraints at hand.

Finally, there are many interesting questions about how to best
approach the development of practical authentication systems based on
these ideas.  These include questions of customized code design and
implementation, but also architectural issues concerning the degree to
these systems can be built from interconnections of existing and often
standardized components --- i.e., existing compression systems,
error-control codes, and public-key cryptographic tools.

\appendix[A Public-Key Adaptation of the Private-Key Authentication
System Model]

To simplify the analysis we have focussed on private key systems where
the encoder and decoder share a secret key $\secKey$, which is kept
hidden from editors.  In most practical applications, however, it is
more convenient to use public key systems where a public key $\pubKey$
is known to all parties (including editors) while a signing key,
$\privKey$, is known only to the encoder.  The advantage of public key
systems is that while only the encoder possessing $\privKey$ can
encode, anyone possessing $\pubKey$ can decode and verify a properly
encoded signal.  In this section, we briefly describe how a secret key
authentication system can be combined with a generic digital signature
scheme to yield a public key system.  Some additional aspects of such
an implementation are discussed in, e.g., \cite{martinian_2001,
mthesis}.

A digital signature scheme consists of a signing function $\dtag =
\dsign(m,\privKey)$ and verifying function $\dver(m,\dtag,\pubKey)$.
Specifically, the signing function maps an arbitrary length message
$m$ to a $\gamma$ bit tag $\dtag$ using the signing key $\privKey$.
The verifying function returns true (with high probability) when given
a message, public key, and tag generated using the signing function
with the corresponding signing key.  Furthermore, it is
computationally infeasible to produce a tag accepted by the verifier
without using the signing key.  Many such digital signature schemes
have been described in the cryptography literature where $\tau$
requires a number of bits that is sub-linear in $n$ or even finite.

\textit{Modified Encoder:}

\begin{enumerate}

\item The public key of the digital signature scheme is published, and
  there is no secret key (equivalently, the secret key in the
  our original formulation is simply published).

\item The encoder uses the original authentication system to map the
source $\ful{\nSrc}$ to $\ful{\tilde{\nChIn}}=\encoder(\ful{\nSrc})$.

\item For a system like the one described in
  \secref{sec:forw-part:-suff}, there are a finite number of possible
  values for the authentic reconstruction $\ful{\nSrch}$ and the
  authentic reconstruction is a deterministic function of
  $\ful{\nSrc}$.  Thus each reconstruction can be assigned a bitwise
  representation $\codeword{ }(\ful{\nSrch})$, from which the encoder
  computes the digital signature tag $\dtag = \dsign(\codeword{
  }(\ful{\nSrch}),\privKey)$ using the digital signature algorithm.

\item Finally the signature $\dtag$ is embedded into
  $\ful{\tilde{\nChIn}}$, producing $\ful{\nChIn}$, using an
  information embedding (data hiding) algorithm.  The chosen algorithm
  can be quite crude since $\dtag$ only requires a sub-linear number
  of bits.  The algorithm parameters are chosen to that the embedding
  incurs asymptotically negligible additional distortion to the overall
  encoding process.

\end{enumerate}

\textit{Modified Decoder:}

\begin{enumerate}

\item The decoder extracts from $\ful{\nChOut}$ an estimate
  $\hat{\dtag}$ of the embedded signature $\dtag$.  Since the size of
  $\dtag$ is sub-linear, the embedding algorithm parameters can be
  further chosen so that $\hat{\dtag} = \dtag$ with arbitrarily high
  probability when the reference channel is in effect.

\item Next, the decoder uses the original authentication system to
  produce $\ful{\nSrct}=\xdecn{\ful{\nChOut}}$, and then, in turn, its
  bitwise representation $\codeword{ }(\ful{\nSrct})$.

\item The decoder checks whether the digital signature verifying
  algorithm $\dver(\codeword{ }(\ful{\nSrct}),\hat{\tau},\pubKey)$
  accepts the $\ful{\nSrct}$ as valid.

\item If so, then the decoder produces the authentic reconstruction
  $\ful{\nSrch}=\ful{\nSrct}$.  Otherwise, the decoder produces the
  special symbol $\dfail$, declaring that it is unable to
  authenticate.

\end{enumerate}

With this construction, we see that the security of such a system is
determined by the security of the underlying public-key digital
signature scheme used.  Specifically, the only way an attacker can
defeat the system is to find a matching $\ful{\nSrch}$ and $\dtag$
accepted by the digital signature verifying algorithm.  All other
performance aspects of the system are effectively unchanged.

\section*{Acknowledgment}

The authors are grateful to Prof.~Ram Zamir for many helpful
suggestions including improvements to the proof of the converse part
of \thrmref{th:main}.  The authors would also like to thank the
reviewers and associate editor for their careful reading of the
manuscript and suggestions for improvement.

%


\begin{biographynophoto}{Emin Martinian}

  (S'00-M'05) completed his undergraduate degree
  in electrical engineering and computer science at the University of
  California at, Berkeley in 1997. After a year and a half at the
  startup OPC Technologies, he joined the doctoral program at MIT in
  1998, receiving the masters degree in 2000, and the doctoral degree
  in 2004. His masters research was in the area of multimedia
  authentication, and his doctoral thesis was in the area of dynamic
  information and constraints in source and channel coding.

  Since completing his doctorate, he has been working on problems of
  video processing, distribution, and compression at Mitsubishi
  Electric Research Laboratories in Cambridge, MA.  His broader
  research interests include digital communications, signal
  processing, information theory, belief propagation, and
  cryptography.  While at MIT he held an NSF Graduate Fellowship, and
  received the Capocelli Award of the 2004 Data Compression Conference
  for the best student-authored paper.

\end{biographynophoto}

\begin{biographynophoto}{Gregory W. Wornell}

  (S'83-M'91-SM'00-F'04) received the
  B.A.Sc.\ degree from the University of British Columbia, Canada, and
  the S.M. and Ph.D. degrees from the Massachusetts Institute of
  Technology, all in electrical engineering and computer science, in
  1985, 1987 and 1991, respectively.

  Since 1991 he has been on the faculty at MIT, where he is Professor
  of Electrical Engineering and Computer Science, co-director of the
  Center for Wireless Networking, and Chair of Graduate Area I
  (Systems, Communication, Control, and Signal Processing) within the
  department's doctoral program.  He has held visiting appointments at
  the former AT\&T Bell Laboratories, Murray Hill, NJ, the University
  of California, Berkeley, CA, and Hewlett-Packard Laboratories, Palo
  Alto, CA.

  His research interests and publications span the areas of signal
  processing, digital communication, and information theory, and
  include algorithms and architectures for wireless and sensor
  networks, broadband systems, and multimedia environments.  He has
  been involved in the Signal Processing and Information Theory
  societies of the IEEE in a variety of capacities, and maintains a
  number of close industrial relationships and activities.  He has won
  a number of awards for both his research and teaching.

\end{biographynophoto}

\begin{biographynophoto}{Brian Chen}

  is a quantitative researcher at the hedge fund Fort Hill
  Capital Management.  He is an alumnus of the Digital Signal
  Processing Group at the Massachusetts Institute of Technology, where
  he received a Ph.D. in Electrical Engineering and Computer Science.
  His areas of expertise include estimation, prediction, and other
  signal processing algorithms, which can be used in such diverse
  applications as financial modeling, multimedia, and communications.
  His Ph.D. thesis explored topics in information hiding and digital
  watermarking.  Some of the techniques described in this thesis were
  exploited by Chinook Communications, a company that he co-founded,
  to alleviate last-mile bandwidth congestion problems in broadband
  networks.

\end{biographynophoto}

\end{document}